\documentclass[fleqn,usenatbib]{mnras}

\usepackage{newtxtext,newtxmath}

\usepackage[T1]{fontenc}

\DeclareRobustCommand{\VAN}[3]{#2}
\let\VANthebibliography\thebibliography
\def\thebibliography{\DeclareRobustCommand{\VAN}[3]{##3}\VANthebibliography}

\usepackage{graphicx}	
\usepackage{xcolor, ulem}
\usepackage{subcaption}
\usepackage{orcidlink}
\usepackage{verbatim}
\usepackage{appendix}

\definecolor{purple_cust}{HTML}{800080}

\graphicspath{ {figures/} }

\title[Kilonova line profiles]{Late-time emission-line profiles from kilonova models}

 \author[Sim, et al.]{
S. A. Sim$^1$\thanks{E-mail: s.sim@qub.ac.uk}\orcidlink{0000-0002-9774-1192}, 
 O. Just$^{2,3}$\orcidlink{0000-0002-3126-9913}, 
 Z. Xiong$^{2}$\orcidlink{0000-0002-2385-6771},
 A. Bauswein$^{2,4}$\orcidlink{0000-0001-6798-3572}, 
 L. P. Mulholland$^{1}$\orcidlink{0009-0003-2668-5589},
 G. Leck$^{2,5}$\orcidlink{0000-0002-0093-0211},
 \newauthor
 G. Martínez-Pinedo$^{2,5,4}$\orcidlink{0000-0002-3825-0131}, 
 L. J. Shingles$^{1,2}$\orcidlink{0000-0002-5738-1612},
 C. E. Collins$^{6}$\orcidlink{0000-0002-0313-7817},
 F. McNeill$^{1}$\orcidlink{0009-0001-9528-7475},
 M. McCann$^{1}$\orcidlink{0000-0002-1532-1240},
 \newauthor
 C. P. Ballance$^1$\orcidlink{0000-0003-1693-1793},
 C. A. Ramsbottom$^{7,1}$\orcidlink{0000-0003-1579-8556}
 \\
 $^{1}$Astrophysics Research Centre, School of Mathematics \& Physics, Queen's University Belfast, BT7 1NN, Northern Ireland.\\
$^{2}$GSI Helmholtzzentrum f\"{u}r Schwerionenforschung, Planckstra{\ss}e 1, D-64291 Darmstadt, Germany.\\
$^{3}$Astrophysical Big Bang Laboratory, RIKEN Cluster for Pioneering Research, 2-1 Hirosawa, Wako, Saitama 351-0198, Japan.\\
$^{4}$Helmholtz Forschungsakademie Hessen f\"{u}r FAIR, GSI Helmholtzzentrum f\"{u}r Schwerionenforschung, Planckstra{\ss}e 1, 64291
Darmstadt, Germany.\\
$^{5}$Institut f\"{u}r Kernphysik (Theoriezentrum), Fachbereich Physik,
Technische Universit\"{a}t Darmstadt, Schlossgartenstra{\ss}e 2, D-64289 Darmstadt, Germany.\\
$^{6}$School of Physics, Trinity College Dublin, The University of Dublin, Dublin 2, Ireland.\\
$^{7}$School of Physics, University College Dublin, Dublin 4, Ireland.
 }

\date{Accepted XXX. Received YYY; in original form ZZZ}

\pubyear{2026}

\begin{document}
\label{firstpage}
\pagerange{\pageref{firstpage}--\pageref{lastpage}}
\maketitle

\begin{abstract}
Numerical simulations suggest that neutron star mergers eject material with complex, non-spherical density and composition distributions.
Here we use two-dimensional configurations of merger ejecta obtained from long-term hydrodynamic simulations to quantify the influence of such ejecta structure on the shapes of spectral lines in the optically thin limit. We consider three example elements of interest for kilonova modelling (selenium, tellurium and tungsten) and illustrate profile shapes for a sample of models and observer orientations. 
Many of our calculations yield complex profile shapes, including cases with multiple peaks and/or extended wings on scales large enough to be relevant to interpreting observations. For selenium and tellurium, our late-phase profile shapes are most sensitive to the structure of the low-velocity ejecta ($\lesssim 0.1\,c$) launched after the merger from the relic black-hole torus system, while for heavier elements the contribution from the more rapidly expanding and more neutron-rich dynamical ejecta launched right after the merger is more significant and leads to broader line shapes. We also find that the dynamical influence of heating due to the decay of r-process elements can lead to considerably broader peaks than suggested by models that neglect this effect. Although idealised, our calculations demonstrate that line shapes are sensitive to the ejecta structure and could therefore constrain the polar observation angle or underlying properties of the merger that determine the spatial distributions of elements in the ejecta components, such as the binary mass ratio or even the equation of state of high-density matter.
\end{abstract}




\section{Introduction}

Multi-dimensional neutrino-hydrodynamics simulations suggest that the distribution of elements in the ejecta from neutron star (NS) mergers can have complex geometries \citep[e.g.][]{Metzger2014, Perego2014a, Wanajo2014a, Just2015a, Foucart2015, Radice2016a, Combi2023b, Schianchi2024a, Musolino2025a, Cheong2025b}. In long-term evolution models (e.g. \citealp{Fernandez2015c, Fujibayashi2020b, just23, Magistrelli2024b, Bernuzzi2025a, Sippens-Groenewegen2025a}) the interplay of different ejecta components typically leads, at high velocities $v\gtrsim 0.1\,c$, to 
regions that are relatively rich in heavy r-process
  (rapid neutron-capture process) elements around the equatorial plane
 (which is dominated by dynamical ejecta with low electron fraction, $Y_e$) and polar regions with lighter elements (higher $Y_e$, primarily originating in outflows from the NS-torus or black-hole (BH) torus merger remnant). The low-velocity domain is typically occupied by ejecta launched from the BH-torus remnant, if formed \citep[e.g.][]{Fernandez2013b, Just2015a, Siegel2017b, Miller2019a, Kawaguchi2025a}. Radiative transfer simulations \citep[e.g.][]{Tanaka2013, Kasen2015, Wollaeger2018a, darbha2020, Heinzel2021e, Kawaguchi2021b, korobkin2021, wollaeger2021,bulla2023,Collins23,shingles23,kawaguchi2024} 
demonstrate that departures from spherical symmetry can affect the light curves and spectra of the kilonova resulting from a NS merger, indicating that ejecta structure is relevant for interpreting observations and testing models. However, modelling early-time spectra is complicated since opacities are high and, moreover, radiative transfer effects can lead to emission that is more symmetric than the underlying distribution of ejecta material \citep{Collins24}. For the only kilonova that has so far been spectroscopically well-observed at early phases \citep[AT2017gfo,][]{abbott17, pian17, smartt17}, empirical modelling finds that the early-time spectrum-forming regions of the ejecta appear consistent with spherical symmetry \citep{sneppen23}. However, further observational data sets will be needed to understand the generality of this finding, and even for AT2017gfo the strength of this constraint diminishes with time, as reducing opacity allows us to see deeper into the ejecta.

In principle, determining the impact of the ejecta structure on observations becomes simpler at later phases.
In particular, if spectral regions become optically thin while homologous expansion is maintained, the ejecta geometry will determine the profile shapes of emission lines in the spectrum \cite[see e.g. ][]{jerkstrand17, jerkstrand25a}. For example, \cite{mccann25} illustrated the impact of different velocity distributions for ejected tungsten (W) on the line-profile shape. Although they considered only spherical ejecta, this highlights that different merger properties 
might impact spectral shapes.
If the ejecta are not spherically symmetric, this will lead to profile shapes that depend on the line of sight, potentially providing diagnostics for the ejecta structure and the observer orientation. 
Most previous studies that consider emission-dominated spectra of kilonovae  \citep{hotokezaka21, hotokezaka23, pognan23, gillanders24, pognan25a, pognan2026, jerkstrand25b,   gillanders2025, ricigliano25,  mccann25} assume either 1D models or adopt a Gaussian line profile shape, meaning that effects of aspherical ejecta and observer orientation are not considered. In this paper, we use outputs from multi-dimensional merger models to explore how line profile shapes might vary for different elements and observer orientations.
Our approach is simplistic and neglects multiple important factors, such as ionization/temperature variations, residual opacity and line blending. But our results illustrate that the multi-dimensional structures predicted by merger simulations can lead to significant variations in profile shapes that should be considered in future modelling
and may provide constraints on the observer orientation and the spatial distribution of the elements.

Throughout this exploratory study, we focus on three selected elements: selenium (Se, $Z=34$), tellurium (Te, $Z=52$) and W ($Z=74$). These are chosen since they lie reasonably close to the three peaks in the r-process nucleosynthesis pattern \citep[see, e.g.,][for reviews on r-process nucleosynthesis]{Arnould2007, Cowan2021g}, and in addition all three have been suggested as candidate identifications associated with observations of late-time emission from kilonovae \cite[e.g. ][]{hotokezaka22, levan2024, gillanders2025}. As such, they provide a good overview of the range of possibilities associated with r-process elements created in the model, and allow for illustrative comparisons between material produced at low $Y_e$ (e.g. W) and products at higher $Y_e$.
The method we use to compute line profile shapes closely follows many previous studies of supernovae and kilonova \citep[see e.g. ][]{jerkstrand17, jerkstrand25a, simotas25, vanBaal24, vanBaal25a, vanBaal25b} and is summarised in Section~\ref{sec-method}. We then overview the merger models used as input for our study in Section~\ref{sec:models} before presenting results in Section~\ref{sec-results} and discussing the implications and outlook in Sections~\ref{sec-discuss} and \ref{sec-outlook}, respectively.

\section{Method}
\label{sec-method}

\subsection{Line shapes in 2D}

We consider 2D ejecta structures with rotational symmetry about the polar ($z$) axis. The ejecta are assumed to be in homologous expansion such that velocity ($\mathbf{v}$) is related to position ($\mathbf{r}$) by $\mathbf{v} = \mathbf{r} / t$, where $t$ is the time since merger. For such cases, the emergent profile for an optically thin emission line (from upper atomic state $u$ to lower state $l$) can be expressed as
\begin{equation}
F_{\Delta \lambda} =\frac{1}{2} \iint n_{u} A_{ul} \eta(\Delta \lambda, \mu) \; r_{c} \mbox{d}r_{c} \; \mbox{d}z
\end{equation}
where $F_{\Delta \lambda}$ is a photon spectrum (photons s$^{-1}$~sr$^{-1}$~\AA$^{-1}$, at wavelength shift $\Delta \lambda$ from line centre), $r_c \equiv \sqrt{x^2 + y^2}$ is the cylindrical radius coordinate,  
$n_{u}$ is the level population (number density) in state $u$ and $A_{ul}$ is the Einstein coefficient for the transition.
$\eta(\Delta \lambda, \mu)$ is a normalised line-profile function that gives the wavelength distribution of photons emitted from a hoop (defined by coordinates $r_c$ and $z$, and rotational symmetry about the $z$-axis), to a distant observer whose line-of-sight unit vector ($\hat{\mathbf{n}}_{\rm obs}$) has direction cosine $\mu \equiv \hat{\mathbf{n}}_{\rm obs} \cdot \hat{\mathbf{z}}$. We consider only first-order Doppler shifts (i.e. $\Delta \lambda = - \lambda_0 \hat{\mathbf{n}}_{\rm obs} \cdot \mathbf{v} /c$, where $\lambda_0$ is the rest wavelength), for which
\begin{equation}
\eta(\Delta \lambda, \mu) = 
\begin{cases}
\frac{ct}{\pi \lambda_0 r_{c} \sqrt{1 - \mu^2}}\frac{1}{\sqrt{1-w^2}}\; , & -1 < w < 1\\
0\; , & \mbox{otherwise}
\end{cases}
\end{equation}
where
\begin{equation}
    w(\Delta\lambda,\mu) \equiv \frac{ct \frac{\Delta \lambda}{\lambda_0} - z\mu}{r_c \sqrt{1-\mu^2}} \; \; .
\end{equation}
The local rate of photon emission (photons s$^{-1}$ cm$^{-3}$) can be expressed as
\begin{equation}
n_{u} A_{ul} = {\rm PEC}_{ul} n_e n_{\rm ion} \label{eqn:pec}
\end{equation}
where $n_{\rm ion}$ is the population of the atom/ion and $n_e$ is the free electron density. The photon-emission-coefficient (${\rm PEC}_{ul}$) depends on the local plasma conditions (i.e. temperature and density, see next sub-section). It is convenient to relate the ion population to the total element population $n_{\rm elem}$ via
\begin{equation}
n_{\rm ion} = f_{\rm ion} n_{\rm elem}
\end{equation}
where $f_{\rm ion}$ is the ion fraction (i.e. the fraction of the element that is in the ionisation state to which the transition of interest belongs). 
For the exploratory investigations here, we make the significant simplification of assuming that the degree of ionization and the temperature are uniform within the ejecta, leading to 
\begin{equation}
F_{\Delta \lambda} \propto 
\int {\rm PEC}_{ul} n_{e} \eta(\Delta \lambda, \mu) \mbox{d} M_{\rm elem} \; 
\end{equation}
where we have converted to an integral over the mass distribution of the element ($M_{\rm elem}$) using
\begin{equation}
2\pi n_{\rm elem} r_{c} \mbox{d}r_{c} \; \mbox{d}z = \frac{1}{m_{\rm elem}} \rho_{\rm elem} \mbox{d}V = \frac{1}{m_{\rm elem}} \mbox{d} M_{\rm elem}
\end{equation}
where $m_{\rm elem}$\footnote{In principle, isotopic variations can mean that $m_{\rm elem}$ is not uniform in the ejecta. However, we assume $m_{\rm elem}$ is constant, and absorb it in the leading proportionality. Neglecting variations in $m_{\rm elem}$ is unlikely to be a significant source of uncertainty compared to our assumption of uniform ionisation.} is the atomic mass, $\rho_{\rm elem}$ is the mass density of the element and $V$ is volume.

\subsection{Density dependence of the photon emission coefficient}
\label{sec:pec}

We focus on optically thin transitions excited by electron collisions, as appropriate for forbidden transitions between low-lying states of the heavy elements. In such cases, at fixed temperature, the ${\rm PEC}$ will tend to a constant value
at low density\footnote{At low density, low-lying excited state level populations are controlled by a competition between collisional excitation and radiative decay that leads to $n_u \propto n_e$, commonly referred to as the {\it coronal regime}. See also discussion by \cite{simotas25}.} but will drop linearly with $n_e$ at high densities.\footnote{At high densities, collisions dominate both excitation and de-excitation, leading to Boltzmann populations.} Accordingly, it is well matched by the simple form

\begin{equation}
{\rm PEC}_{ul} \propto 
\frac{n_{\rm crit}}{n_e + n_{\rm crit}} \, ,
\label{eqn:critical}
\end{equation}
where $n_{\rm crit}$ is the critical density for the transition (see Appendix~A). Under our assumption of uniform ionization, and further assuming that the mean particle mass is approximately uniform, we can take $n_{e}$ as proportional to the total mass density ($n_{e} \propto \rho$) and recast the critical free-electron density as an equivalent critical total mass density ($\rho_{\rm crit}$) such that  

\begin{equation}
F_{\Delta \lambda} \propto \int 
\frac{\rho_{\rm crit} \rho}{\rho + \rho_{\rm crit}} \eta(\Delta \lambda, \mu) \; \mbox{d} M_{\rm elem} \, .
\label{eqn:profile_mass}
\end{equation}

\subsection{Construction from tracer particles}

Eqn. \ref{eqn:profile_mass} provides an easy way to compute line profile shapes for any element mass distribution that has cylindrical symmetry. We apply this to simulations carried out by, or following from, \citet[][hereafter J23]{just23}, making use of the Lagrangian tracer particles that are used for the nucleosynthesis post-processing calculations (as detailed in that work). Each such particle represents a specified mass element with detailed composition. Thus, the mass integral is replaced by a sum over all the particles (indexed by $i$):

\begin{equation}
F_{\Delta \lambda} \propto \sum_i 
\frac{\rho_{\rm crit} \rho}{\rho + \rho_{\rm crit}} \eta(\Delta \lambda, \mu) M_{{\rm elem},i}
\label{eqn:particle_summation}
\end{equation}
where $\eta(\Delta \lambda, \mu)$ is evaluated for each particle at its $r_c$ and $z$ coordinates. To evaluate the density pre-factor we construct  $\rho$ by binning the particles on a 2D spherical polar grid and computing densities from their summed masses and grid cell volumes. We treat $\rho_{\rm crit}$ as a parameter in our calculations and discuss its influence below. Note that in the limit $\rho\gg \rho_{\rm crit}$ the density pre-factor evaluates to $\rho_{\rm crit}$ (i.e. to a constant) whereas for $\rho\ll \rho_{\rm crit}$ it becomes $\rho$, meaning that at low densities the emissivity is additionally weighted by the density. In all cases we take tracer particle compositions and positions at $t = 1$ month ($= 30$~days) post merger.

\subsection{Wavelength binning}

We calculate line profiles numerically by summing contributions to wavelength bins from all tracer particles. Specifically, we compute $F_{\Delta\lambda}$ on a uniformly binned wavelength grid (of wavelength bin size $\delta \lambda$) via
\begin{equation}
F_{\Delta\lambda_j} \equiv \frac{1}{\delta \lambda} \int_{\Delta\lambda_j - \delta\lambda/2}^{\Delta\lambda_j + \delta\lambda/2} F_{\Delta\lambda} \; \mbox{d}\Delta\lambda
\end{equation}
We note that the form of $\eta$ can be readily integrated to find $\int F_{\Delta \lambda} \; \mbox{d}\Delta\lambda$:

\begin{equation}
\int_{\Delta\lambda_{\rm min}}^{\Delta\lambda_{\rm max}}\eta(\Delta \lambda, \mu) 
\; \mbox{d} \Delta\lambda = 
 - \frac{1}{\pi} \left[ {} \cos^{-1} w\right]_{\min(w(\Delta\lambda_{\rm min},\mu), -1)}^{\max(w(\Delta\lambda_{\rm max},\mu),+1)}
\end{equation}

\begin{table*}
  \caption{For each ejecta model included in this study we give the sum of gravitational masses of the NS binary system, the mass ratio, whether r-process heating was included in the hydrodynamic simulation, the lifetime of the NS remnant until BH formation, the masses of the three ejecta components (dynamical, NS-torus, BH-torus) and their sum (tot), and the ejected masses of selected elements. As we assume equatorial symmetry in this study, all masses are measured in the northern hemisphere ($0<\theta<\pi/2$) and multiplied by a factor of two. Models sym-long and sym-heat-long are taken from \protect\cite{Just2025b}, models sym-short and asy-short from \protect\cite{sneppen2026}, model asy-long is similar to model asy-n1-a6 from J23 but with small modifications (which are summarized in footnote 8 of \protect\citealt{Just2025b}), and the three corresponding models sym-heat-short, asy-heat-short, and asy-heat-long use the same setup but with nuclear r-process heating taken into account (using the RHINE scheme described by \protect\citealt{Just2025b}).
  } \label{tab:models}
\begin{tabular}{ccccccccc} \hline
Ejecta model  & $M_1+M_2$     & $M_1/M_2$  & Nucl. heating & $t_{\rm BH}$  & Ejecta masses                  & \multicolumn{3}{c}{Element masses [$M_{\odot}$]}                     \\
              & [$M_\odot$]   &            &               & [ms]          & dyn/NS/BH{/tot} [$10^{-2} M_{\odot}$]   & Se                    & Te                   & W                     \\ \hline
sym-long      & 2.75          & 1          & no            & 214           & 0.51/4.16/4.81{/9.48}                    & $2.4 \times 10^{-2}$  & $5.8 \times 10^{-3}$ & $3.9 \times 10^{-5}$  \\
sym-heat-long & 2.75          & 1          & yes           & 214           & 0.51/4.16/5.99{/10.66}                    & $2.5 \times 10^{-2}$  & $9.7 \times 10^{-3}$ & $3.8 \times 10^{-5}$  \\ \hline
asy-long      & 2.75          & 0.75       & no            & 98            & 1.20/1.98/6.63{/9.81}                    & $2.1 \times 10^{-2}$ & $8.3 \times 10^{-3}$ & $ 7.2 \times 10^{-5}$ \\
asy-heat-long & 2.75          & 0.75       & yes           & 98            & 1.22/1.94/6.97{/10.13}                    &  $1.9 \times 10^{-2}$&  $1.2 \times 10^{-2}$& $7.7 \times 10^{-5}$ \\ \hline
sym-short     & 2.8           & 1          & no            & 10            & 0.76/0.00/0.44{/1.20}                    & $ 1.3 \times 10^{-3}$ & $8.4 \times 10^{-4}$ & $ 5.0 \times 10^{-5}$ \\ 
sym-heat-short& 2.8           & 1          & yes           & 10            & 0.76/0.00/0.50{/1.26}                    & $ 1.5 \times 10^{-3}$ & $ 9.6 \times 10^{-4}$ & $ 5.1 \times 10^{-5}$ \\ \hline
asy-short     & 2.8           & 0.75       & no            & 10            & 1.19/0.00/1.28{/2.47}                    & $3.6 \times 10^{-3}$  & $1.7 \times 10^{-3}$ & $8.3 \times 10^{-5}$  \\ 
asy-heat-short& 2.8           & 0.75       & yes           & 10            & 1.12/0.00/1.48{/2.60}                    & $3.6 \times 10^{-3}$& $2.4 \times 10^{-3}$ & $9.1 \times 10^{-5}$\\ \hline
\end{tabular}
\end{table*}

\section{Ejecta structures}
\label{sec:models}

We will consider eight different ejecta structures. These are drawn from NS merger simulations (performed with the methodology of J23), and some key properties (total ejecta mass and selected ejected elemental masses) are summarised in Table~\ref{tab:models}. 
The models include example geometries from both symmetric and asymmetric binaries\footnote{That is, binaries with equal or unequal mass components.} that involve either short-lived ($\sim$ a few ms) or long-lived ($\gtrsim$ 100\,ms) massive post-merger NS remnants (prior to eventual collapse to a BH at $t=t_{\rm BH}$). For each setup investigated, we consider a pair of simulations, one in which nuclear heating through the release of radioactive decay energy of synthesised $r$-process nuclei is included in the hydrodynamical evolution (i.e. using the RHINE scheme described by \citealt{Just2025b}, denoted here as `heat' models), and one in which it is not.
Our set of models is not exhaustive in terms of the full range of possible ejecta morphologies, but it is illustrative of possible outcomes grounded in current long-term simulation models that capture both early and secular ejecta components. As such, it provides an interesting set for exploring possible variations of line profile shapes with observer orientation.

The merger simulations were performed in 3D until 10\,ms post merger and then continued with an axisymmetric evolution code until 100\,s, at which time the ejecta configurations used here are extracted (see J23 for more details). Hence, the ejecta models are all 2D, having rotational symmetry about the $z$-axis. 
For the calculations in the body of this paper we also assume reflection symmetry in the $xy$-plane, adopting ejecta structures from the positive-$z$ hemisphere (i.e. $0 \leq \theta \leq \pi/2$, where $\theta$ is the polar angle) of each simulation as input.
However, in Appendix~B we discuss how results are affected if we consider the full domain ($0 \leq \theta \leq \pi$) for two models.

 As for the 2D polar grid used to sample the tracers, the angular grid is uniform and consists of 40 zones between $0 \leq \theta \leq \pi/2$, and the velocity grid spans from $0.005\,c$ to $0.5\,c$ with a zone width $\Delta v$ increasing non-linearly with velocity. The first three of our models in Table~\ref{tab:models}) use $N_v=50$ velocity zones with $\Delta v$ increasing from $0.0008\,c$ to $0.03\,c$, while the remaining models adopt $N_v=100$ zones with $\Delta v=0.0004\,c\ldots 0.015\,c$. Both resolutions should be sufficient for our purpose of examining the basic line-shape features. The lower limit (at $0.005\,c$) is necessary to reduce artifacts from material expanding with very low velocities that may not be resolved reliably. This is because to follow the ejected material for a long timescale with reasonable computational demand, the hydrodynamic simulations after $t=10\,$s remove the small amount of remaining disk material at radii $r<10^9\,$cm {(typically corresponding to less than 3\,\% of the total ejecta mass)} and place a reflecting inner boundary at $r=10^9\,$cm (as described in J23). This treatment was found to overestimate\footnote{One could instead use an open boundary at $r=10^9\,$cm, however, in this case the low-velocity masses are likely to be underestimated instead. A more accurate treatment would thus call for a longer evolution of the inner ($r<10^9\,$cm) domain, which however would make the simulations significantly more expensive. Notably, this source of error is unlikely to play a role in models including r-process heating, in which the very low velocity regime is barely populated. 
{For models that do not include r-process heating, ignoring the contribution from the region $v<0.005\,c$ may possibly underestimate the strength of emission in the line core, depending on whether material at such low velocities can significantly contribute. However, considering that the mass-velocity distribution of the disk ejecta typically peaks at significantly larger values than $0.005\,c$ \citep[e.g. ][]{Fernandez2024a}, the impact is still likely to be small.}}
  the ejecta masses (and densities) at $v \lesssim0.005\,c$, which is why we ignore the data at such low velocities.

\begin{figure*}
\hspace{1.3cm}\includegraphics[height=0.179\textwidth]{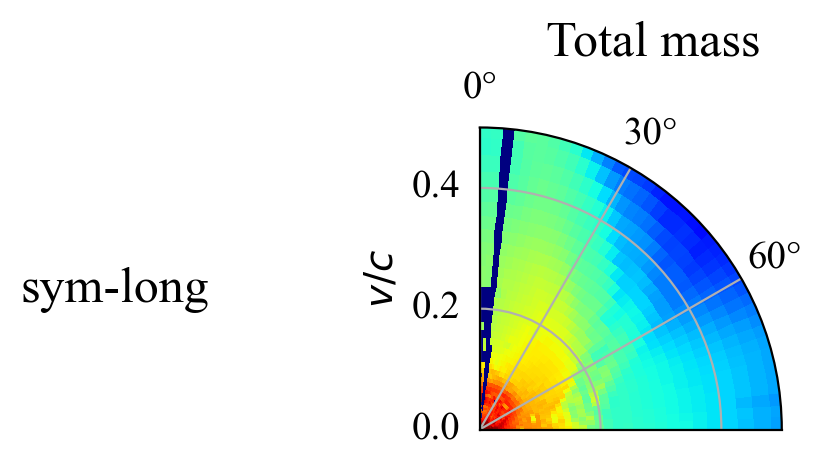}
\includegraphics[height=0.179\textwidth]{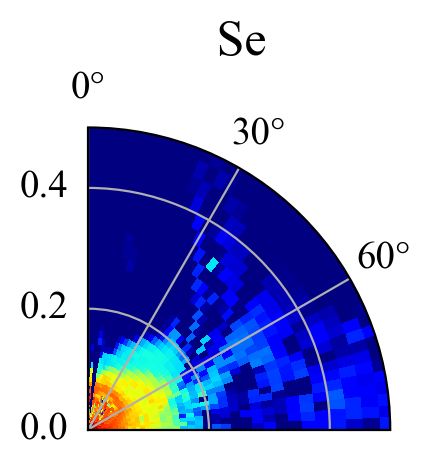}
\includegraphics[height=0.179\textwidth]{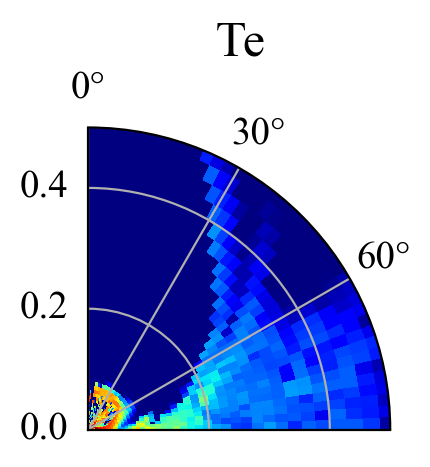}
\includegraphics[height=0.179\textwidth]{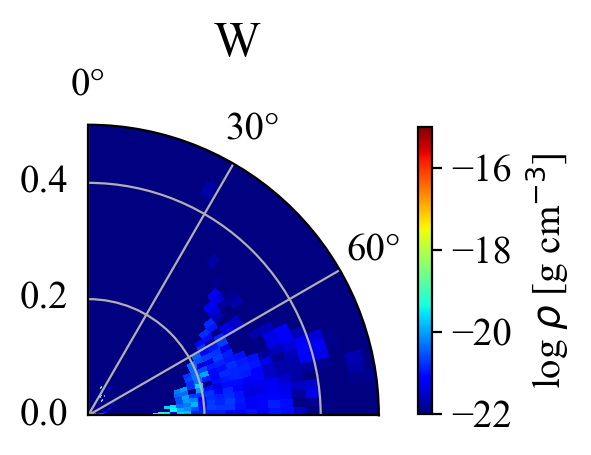}\\ \vspace{-1mm}
\includegraphics[height=0.16\textwidth]{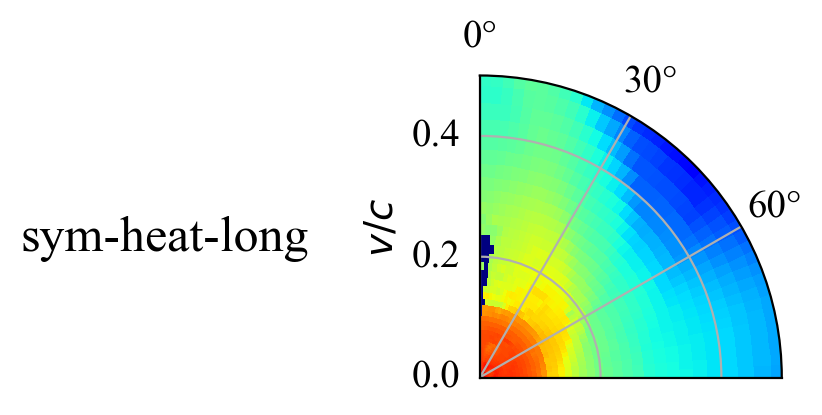}
\includegraphics[height=0.16\textwidth]{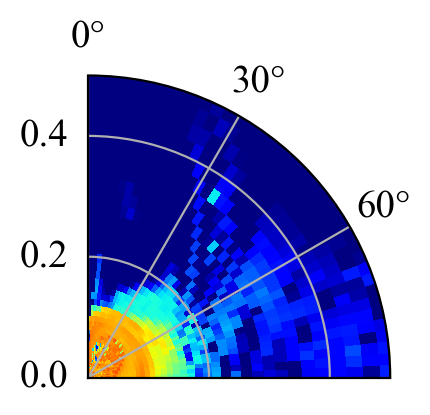}
\includegraphics[height=0.16\textwidth]{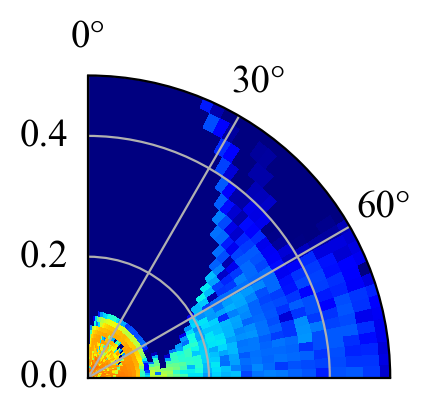}
\includegraphics[height=0.16\textwidth]{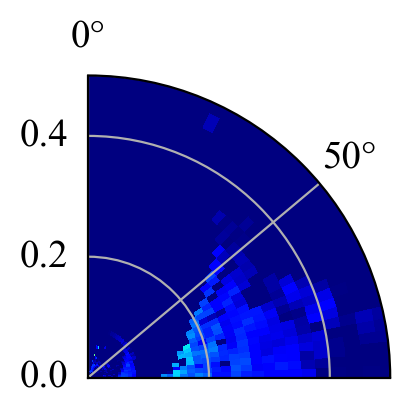}\\ \vspace{-1mm}
\includegraphics[height=0.16\textwidth]{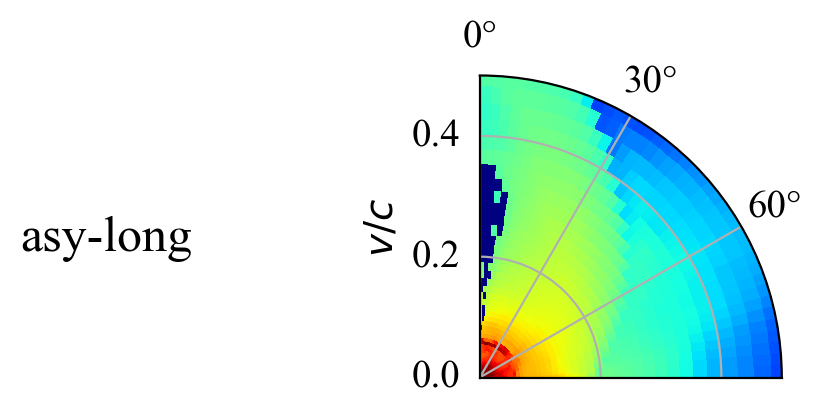}
\includegraphics[height=0.16\textwidth]{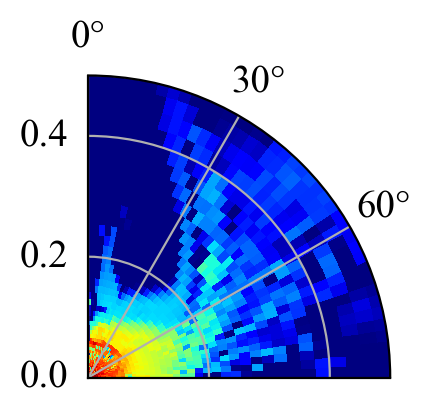}
\includegraphics[height=0.16\textwidth]{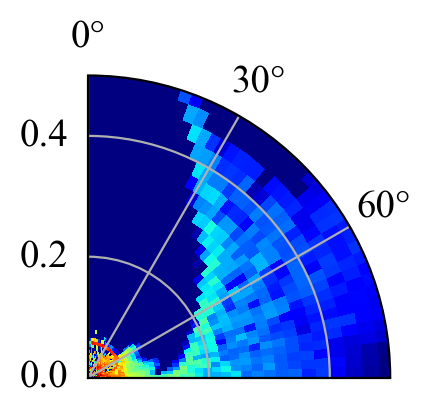}
\includegraphics[height=0.16\textwidth]{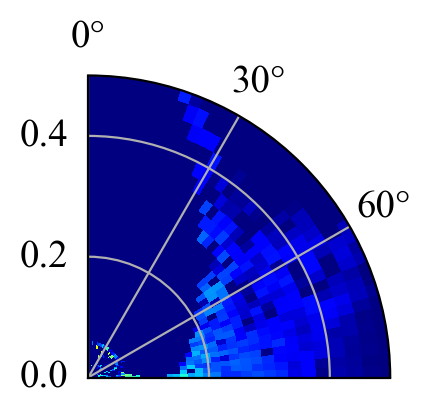}\\ \vspace{-1mm}
\includegraphics[height=0.16\textwidth]{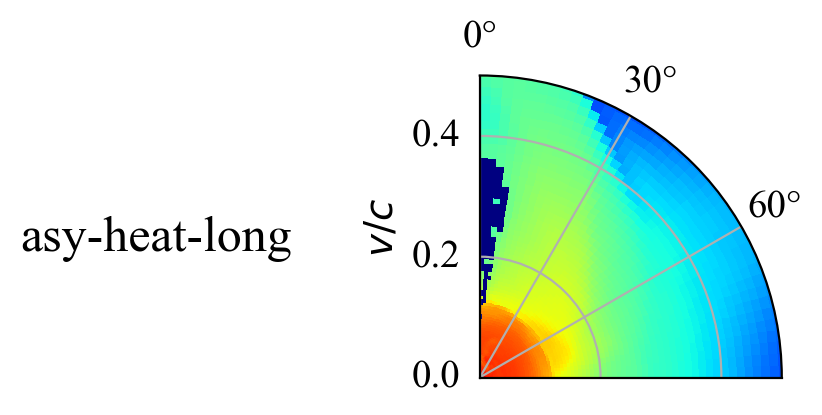}
\includegraphics[height=0.16\textwidth]{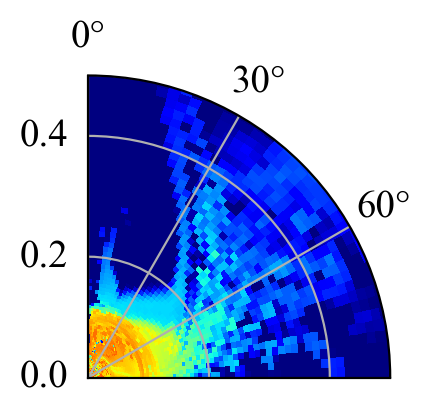}
\includegraphics[height=0.16\textwidth]{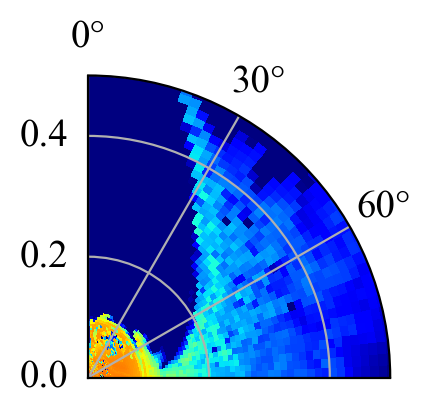}
\includegraphics[height=0.16\textwidth]{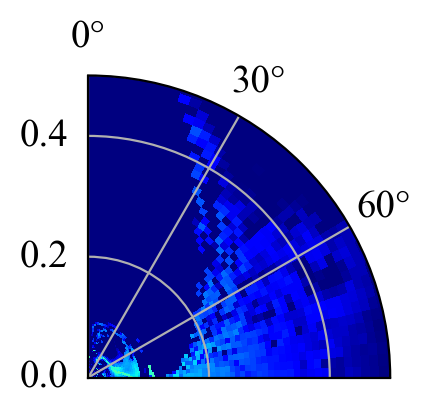}\\ \vspace{-1mm}
\includegraphics[height=0.16\textwidth]{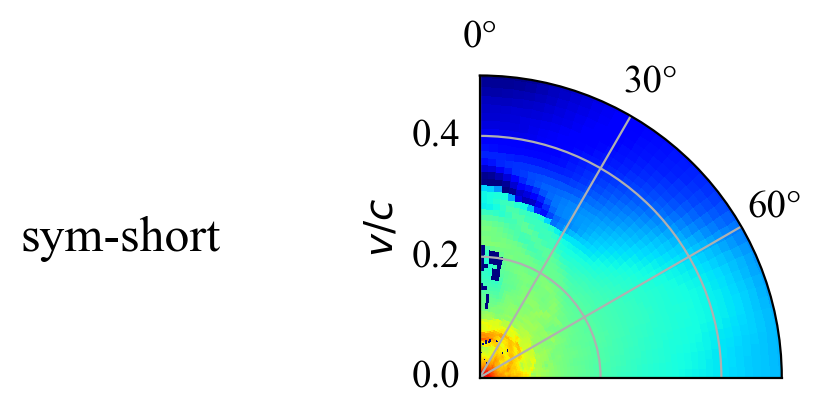}
\includegraphics[height=0.16\textwidth]{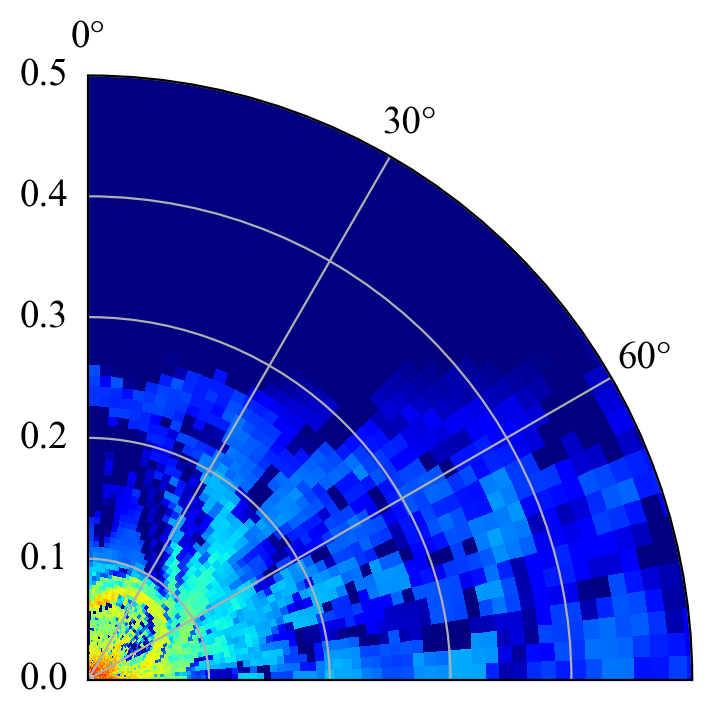}
\includegraphics[height=0.16\textwidth]{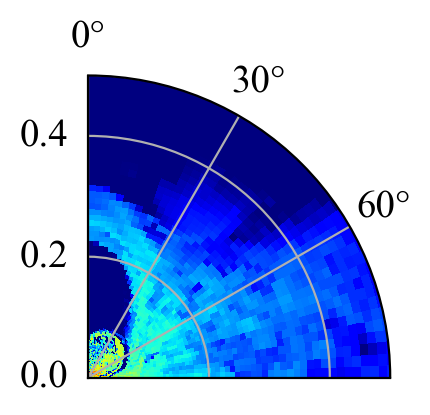}
\includegraphics[height=0.16\textwidth]{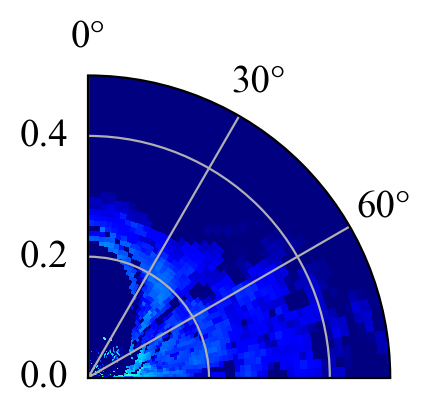}\\ \vspace{-1mm}
\includegraphics[height=0.16\textwidth]{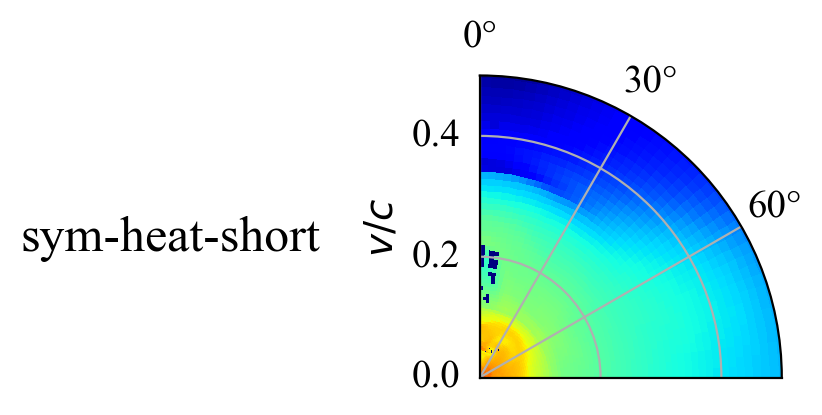}
\includegraphics[height=0.16\textwidth]{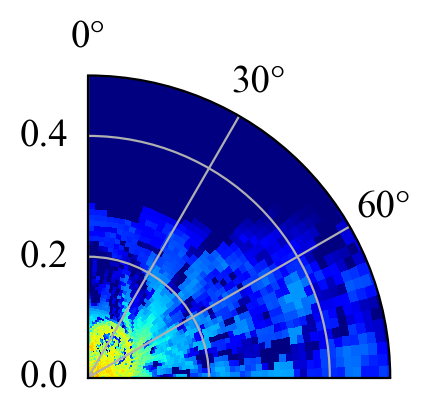}
\includegraphics[height=0.16\textwidth]{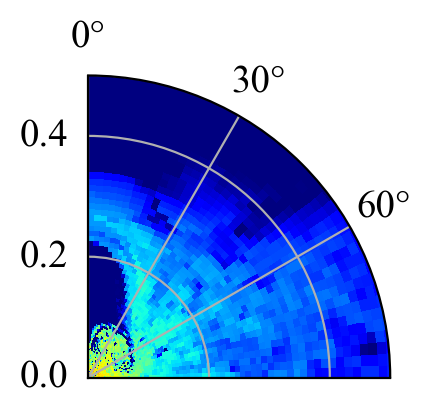}
\includegraphics[height=0.16\textwidth]{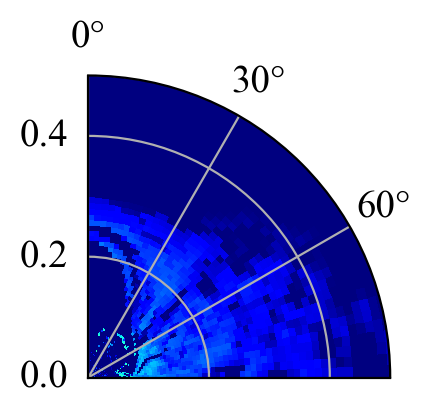}\\ \vspace{-1mm}
\includegraphics[height=0.16\textwidth]{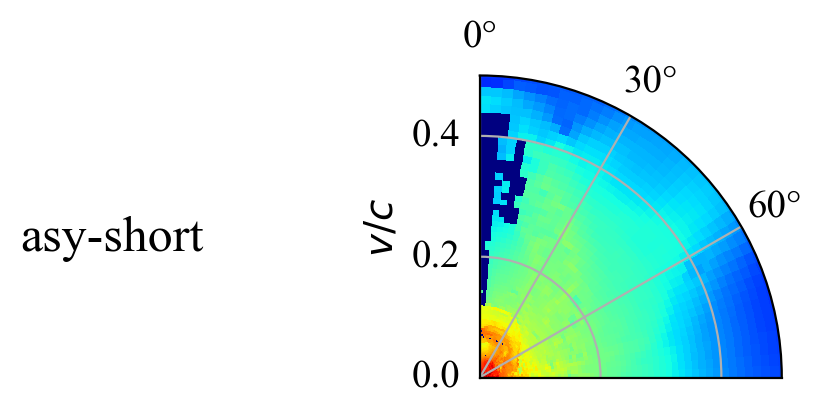}
\includegraphics[height=0.16\textwidth]{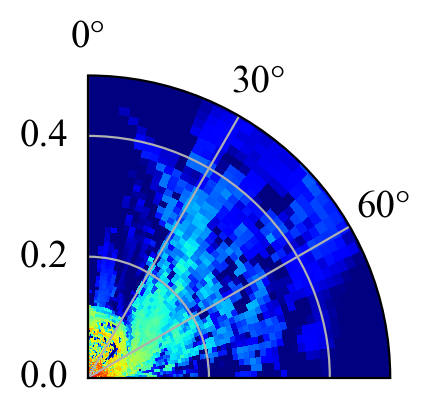}
\includegraphics[height=0.16\textwidth]{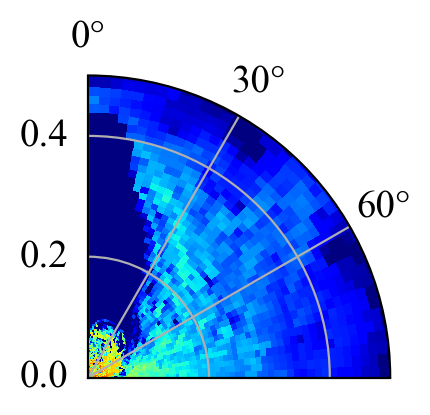}
\includegraphics[height=0.16\textwidth]{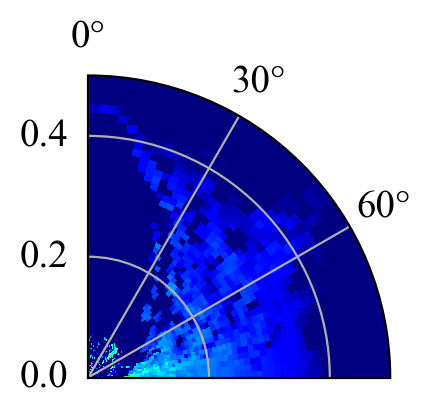}\\ \vspace{-1mm}
\includegraphics[height=0.16\textwidth]{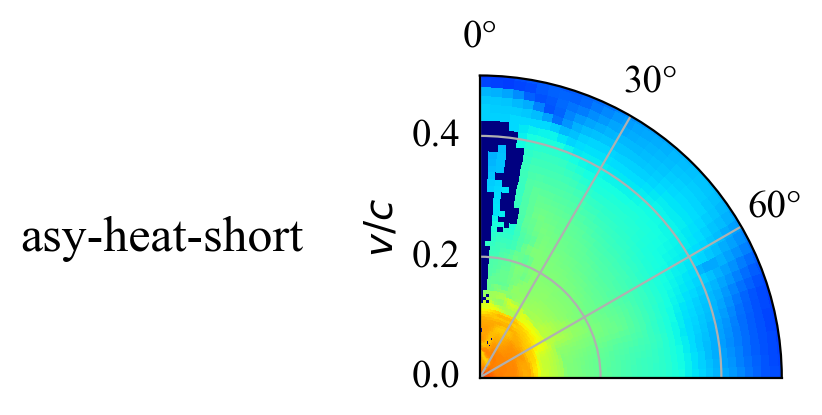}
\includegraphics[height=0.16\textwidth]{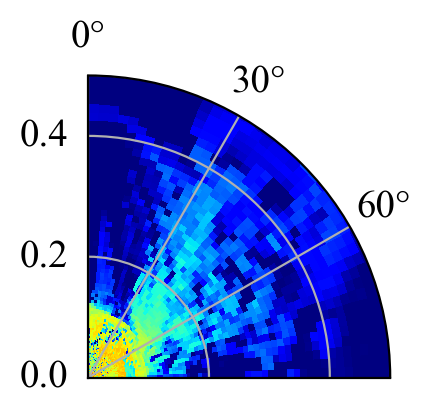}
\includegraphics[height=0.16\textwidth]{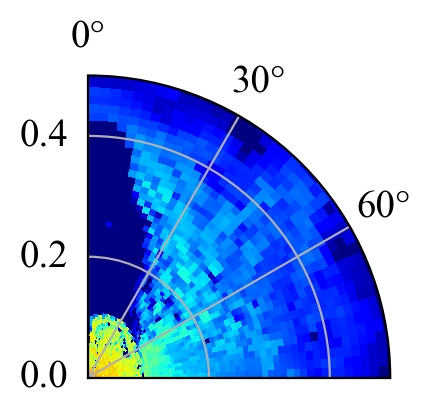}
\includegraphics[height=0.16\textwidth]{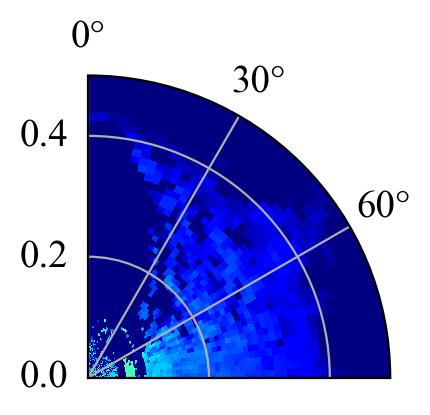}
\caption{Total mass density ($\rho$, left {column}) and elemental mass densities for Se ($Z=34$, 2nd column), Te ($Z=52$, 3rd column) and W ($Z=74$, right column) for models sym-long, sym-heat-long, asy-long, asy-heat-long, sym-short, sym-heat-short, asy-short and asy-heat-short (ordered from top to bottom row). All models are shown in the $x-z$ plane and have rotational symmetry about the $z$ (vertical) axis. Composition is taken at 1 month ($= 30$~days) post merger and, for convenience, densities are plotted in velocity space ($v/c$), since the ejecta are in homologous expansion by this time.}
\label{fig:all_models}
\end{figure*}

The model structures are illustrated in Fig.~\ref{fig:all_models}. {Grid cells with effectively vanishing density close to the polar axis in Fig.~\ref{fig:all_models} are ignored in our analysis and result from trajectories where the path integration backwards in time was unreliable (due to a combination of high fluid velocities and coarse time resolution of available output data). However, the contribution of the corresponding material to the emission lines is presumably insignificant because of the small amount of volume and, therefore, mass sitting in these polar cells}. 

In general, the models have several key features in common. All have their highest densities at low velocity ($\lesssim 0.1$c) but there is significant variation in this region: the models that include $r$-process heating have a more extended inner dense region which also tends to be more spherical than in our other models. As illustrated by the element densities, most of the models have a polar region that is relatively devoid of heavy elements. For example, Te (an element in the second r-process peak, shown in Fig.~\ref{fig:all_models}) tends to be located mainly in the inner ejecta ($\lesssim 0.1$c) or concentrated at (and below) a fairly well defined boundary region that sweeps up from the mid-plane and extends into the outer, high-velocity ejecta for models with long-lived NS remnants (top {four} rows in Fig.~\ref{fig:all_models}). For the short-lived models this outer structure in Te varies, appearing to close back towards the rotation axis. Se (part of the first r-process peak) generally has a qualitatively similar distribution to Te (see second column of Fig.~\ref{fig:all_models}) but extends over a wider range in the inner ejecta and has less concentration around the equatorial plane. In contrast, W (close to the third r-process peak but slightly lighter) is hardly present in the innermost ejecta and is instead concentrated in a similar region to where Te sweeps up from the equatorial plane. 

\begin{figure*}
\includegraphics[width=0.95\columnwidth,height=0.23\textheight,trim={0 150 0 50},clip]{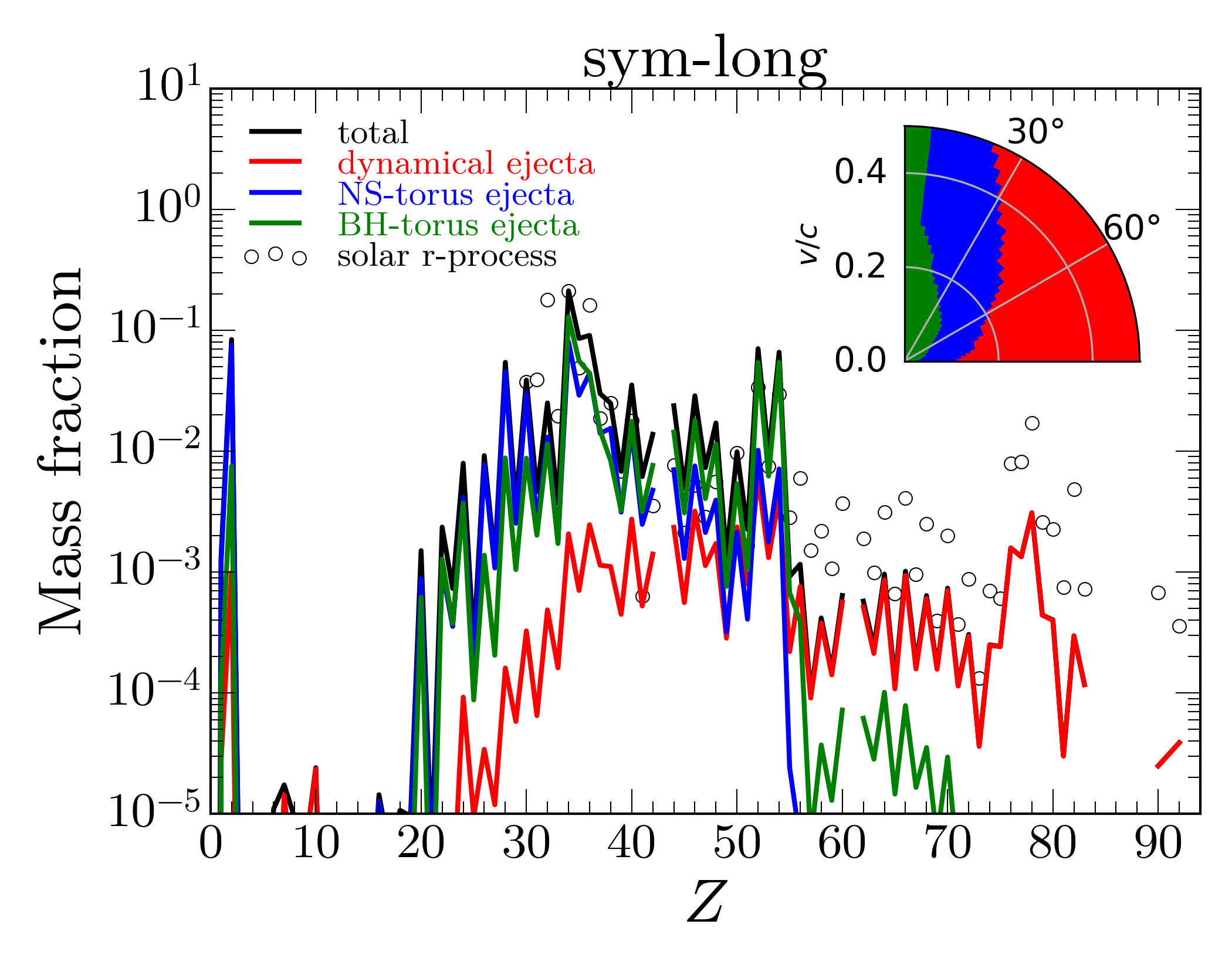}
\includegraphics[width=0.95\columnwidth,height=0.23\textheight,trim={0 150 0 50},clip]{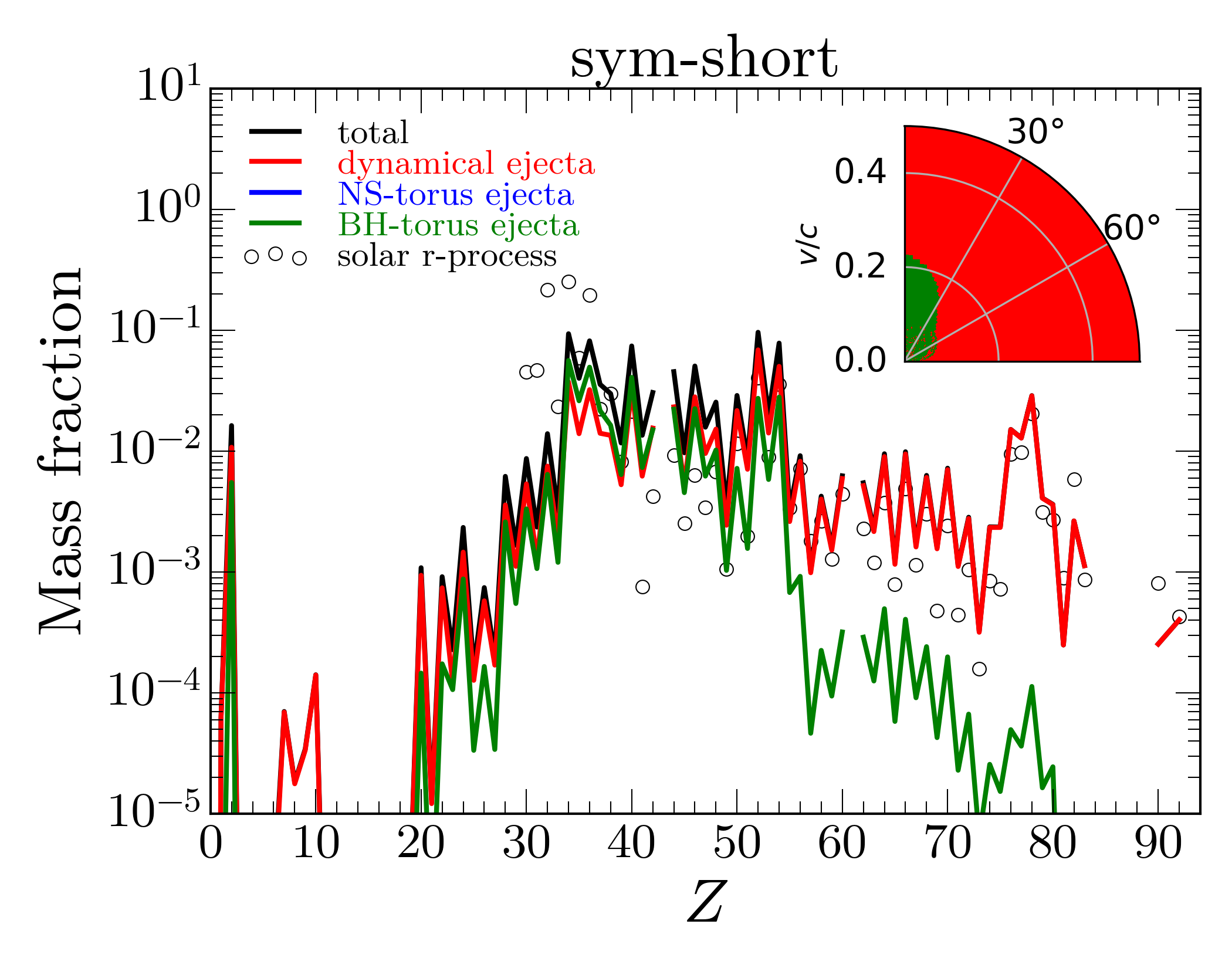}\\[-.5mm]
\includegraphics[width=0.95\columnwidth,height=0.23\textheight,trim={0 150 0 50},clip]{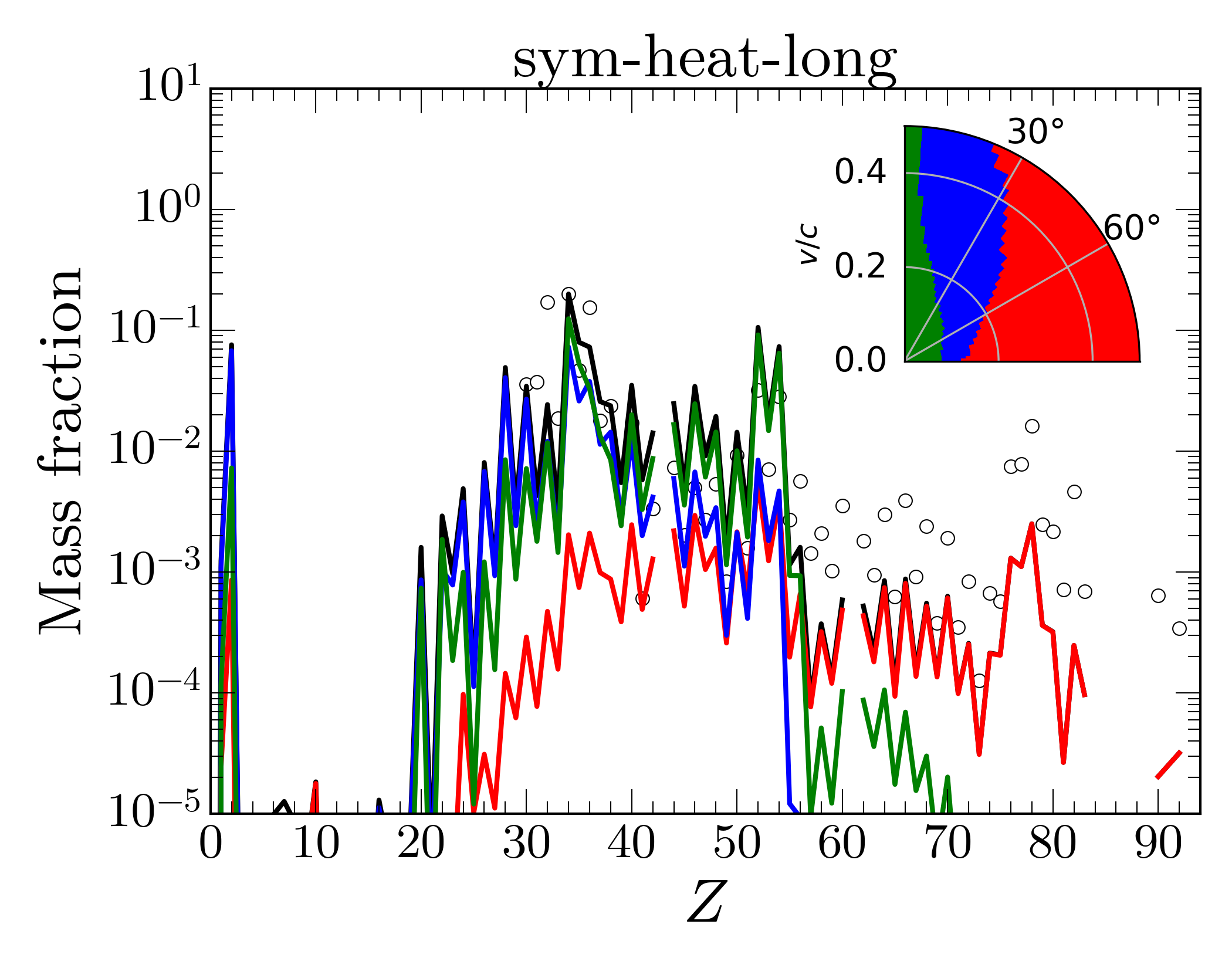}
\includegraphics[width=0.95\columnwidth,height=0.23\textheight,trim={0 150 0 50},clip]{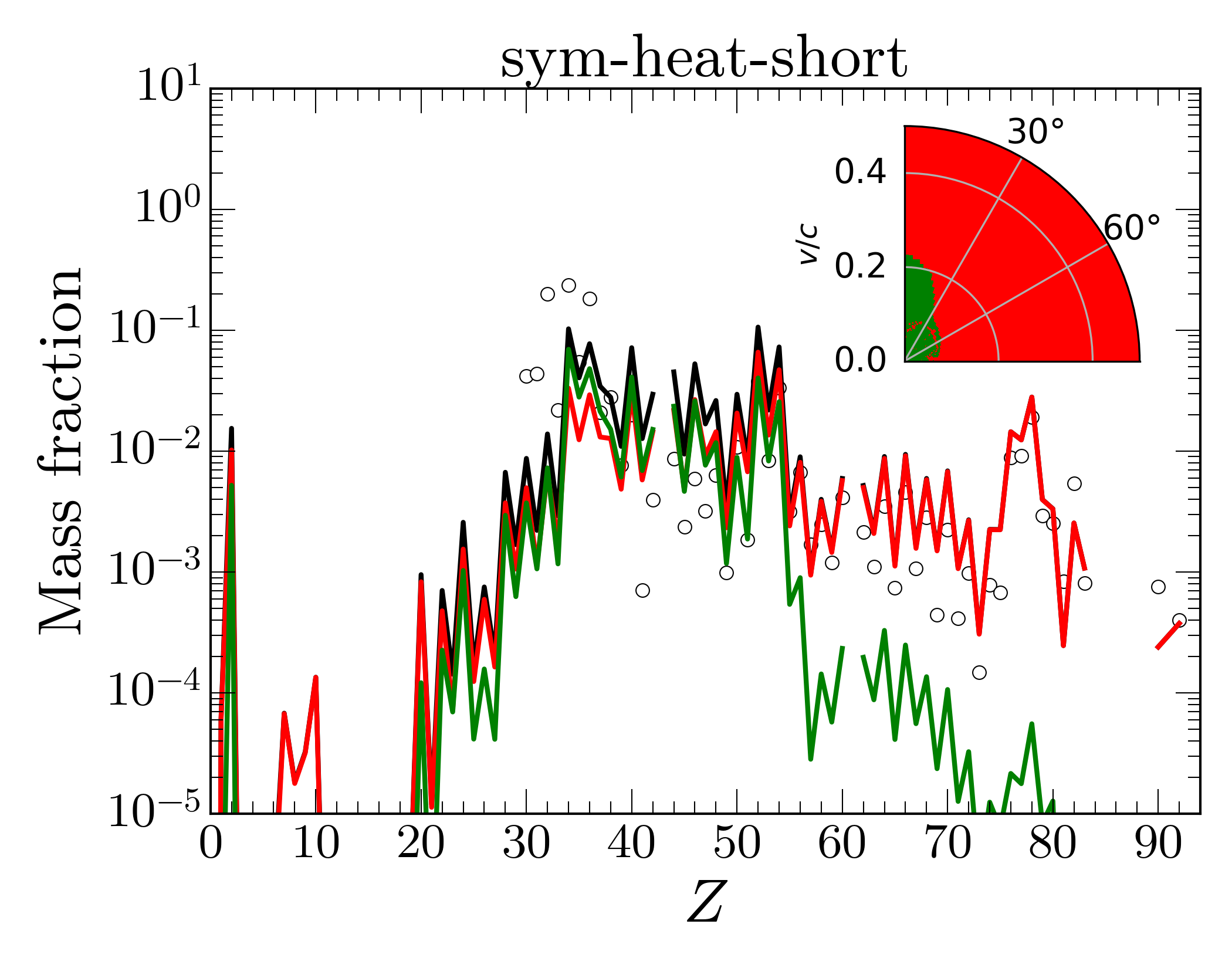}\\[-.5mm]
\includegraphics[width=0.95\columnwidth,height=0.23\textheight,trim={0 150 0 50},clip]{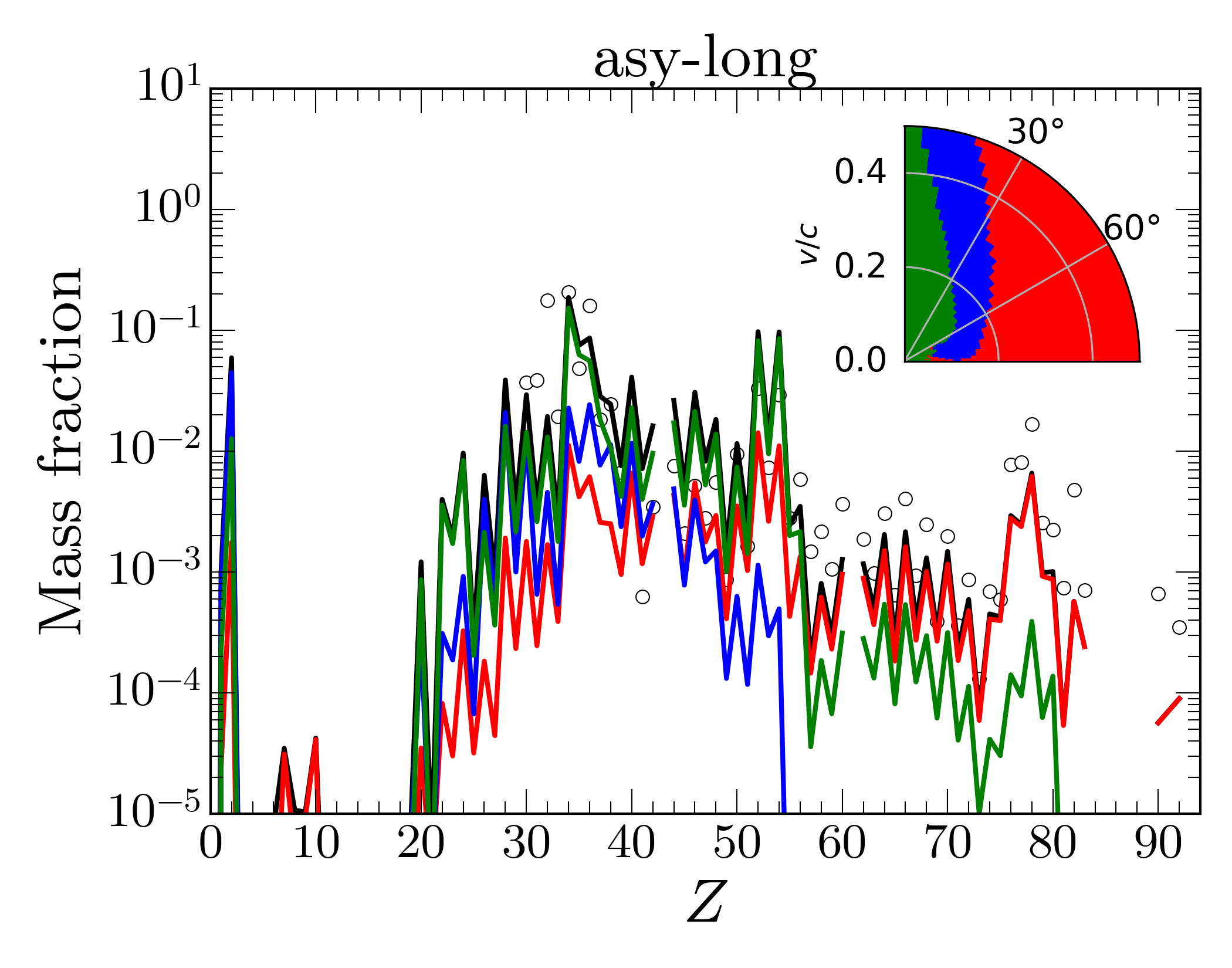}
\includegraphics[width=0.95\columnwidth,height=0.23\textheight,trim={0 150 0 50},clip]{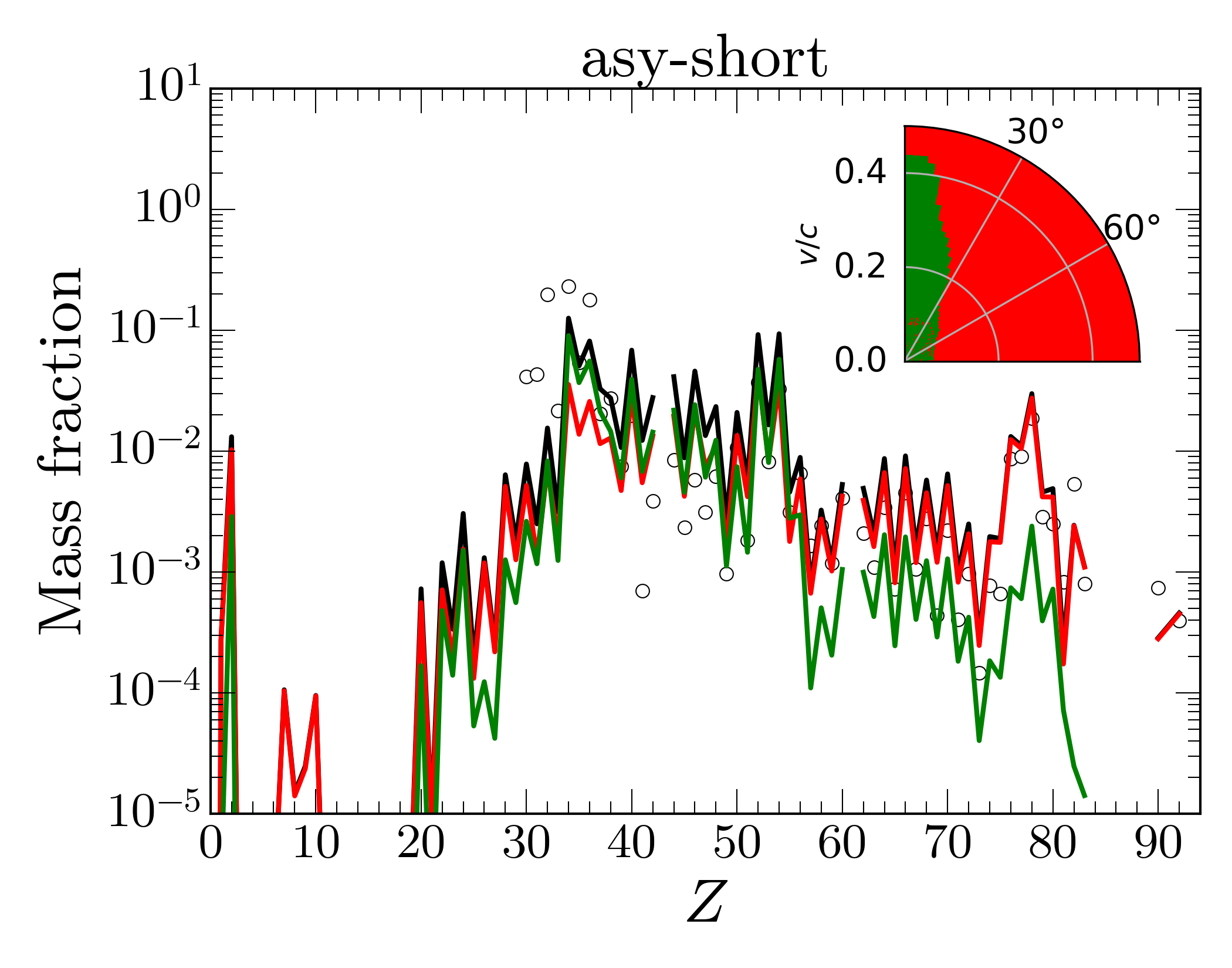}\\[-.5mm]
\includegraphics[width=0.95\columnwidth,height=0.23\textheight,trim={0 0 0 50},clip]{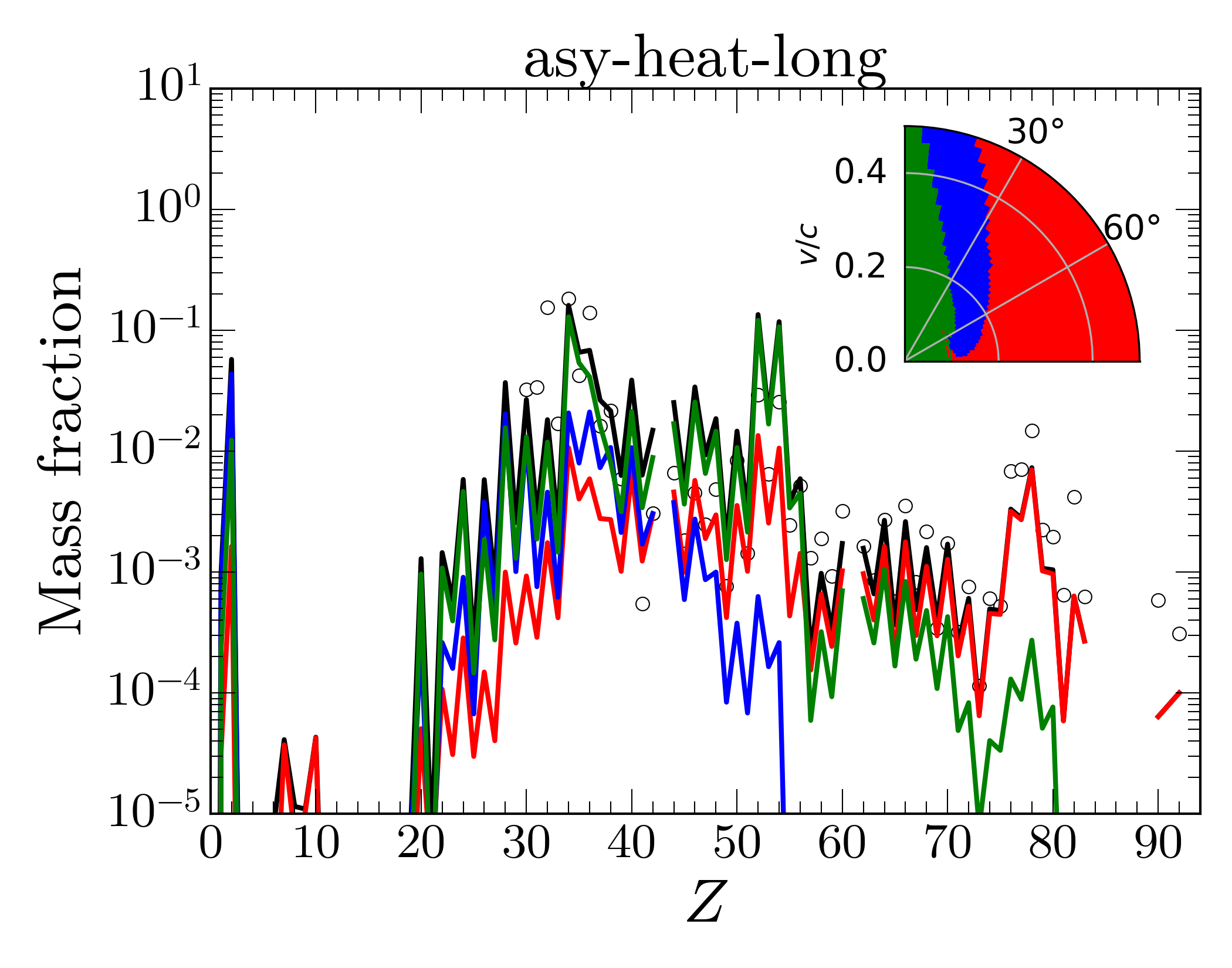}
\includegraphics[width=0.95\columnwidth,height=0.23\textheight,trim={0 0 0 50},clip]{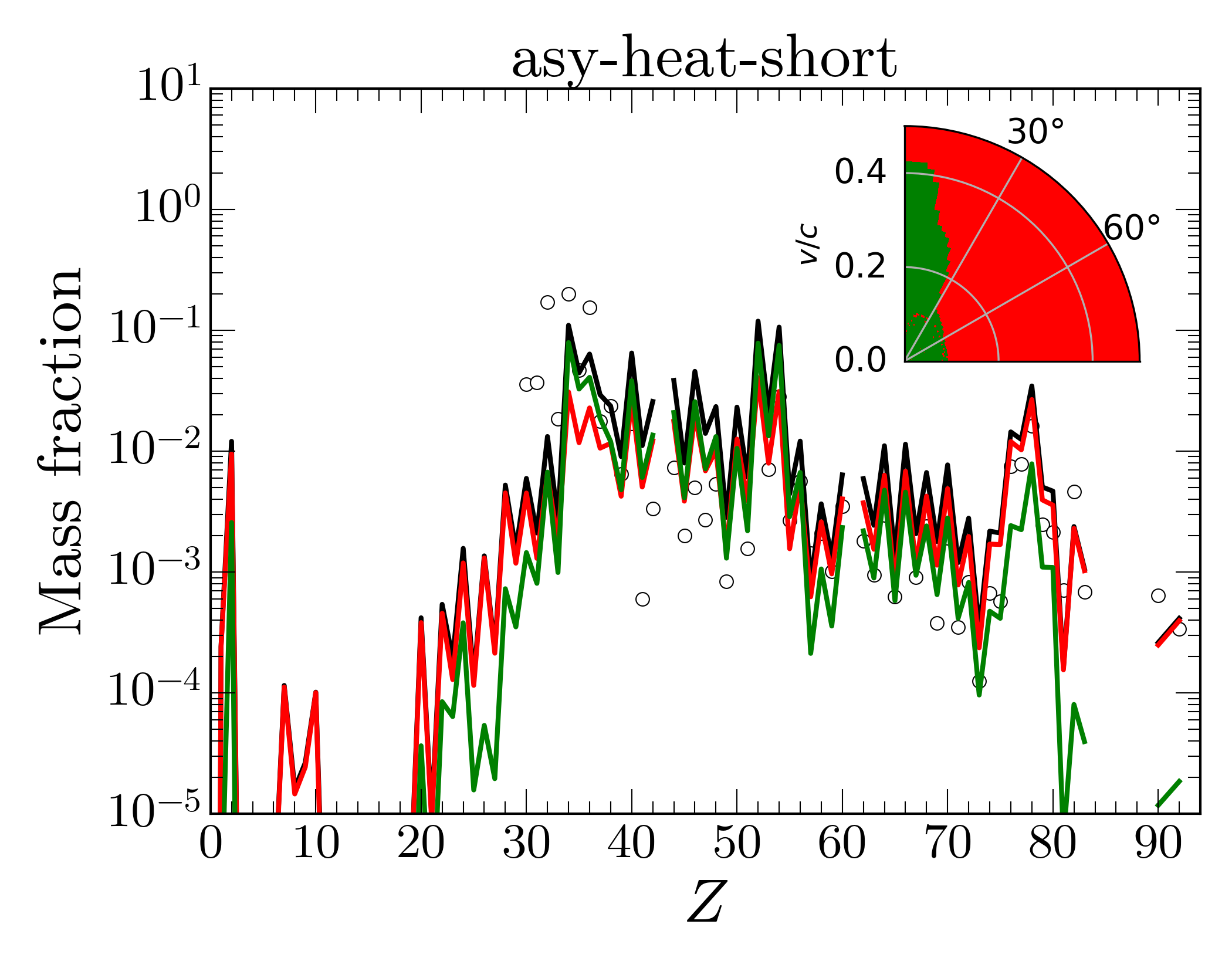}\\[-4mm]
\caption{
{Total nucleosynthesis yields in each ejecta component measured at 1\,month after merger for all models considered in this study. Circles denote the solar r-process abundances \citep{Goriely1999a}. The insets indicate for a given location in velocity space the ejecta component that a tracer belongs to (using the same identification criteria as J23). Heavier nuclei (with atomic numbers $Z\gtrsim 55$) are primarily synthesised in dynamical ejecta, while NS-torus and BH-torus ejecta are dominated by lighter elements. Asymmetric binaries lead to somewhat more neutron-rich BH-tori and, consequently, to a greater amount of heavier elements in the BH-torus ejecta.}}
\label{fig:ej_components1}
\end{figure*}

To better grasp the origin and composition of these ejecta structures, Fig.~\ref{fig:ej_components1} provides information about whether material at a given location was ejected during the merger (denoted as ``dynamical ejecta'', red), during the subsequent evolution of the NS remnant (``NS-torus ejecta'', blue), or during the final disintegration of the BH-torus system (``BH-torus ejecta'', green), along with the nucleosynthesis yields of each ejecta component. As a general trend, dynamical ejecta are on average more neutron rich (and therefore more enriched with heavy elements) than both the NS-torus and BH-torus ejecta owing to their faster expansion and less significant exposure to neutrino irradiation. Overall higher mass fractions of heavy elements ($Z\gtrsim 55$) are found for merger models with a shorter NS-remnant lifetime and with asymmetric binary masses. The wide polar funnel observed for the long-lived models (cf. Fig.~\ref{fig:all_models}) is mainly a result of (less neutron-rich and therefore heavy-element poor) neutrino-driven winds from the NS remnant pushing aside the dynamical ejecta originally escaping along the polar direction. Since in the short-lived models winds from the NS are absent, the polar funnel is less extended (although still present), because the BH-torus system launches only relatively weak polar neutrino winds. The dominant component of the ejecta expanding with $v<0.1\,c$ stems from the BH-torus system and is launched as a result of heating and angular momentum transport caused by turbulent viscosity, which inflates the torus once neutrino cooling becomes inefficient. As a result of the turbulent flow pattern created in the viscous ejecta, the (total and elemental) density profiles (cf. Fig.~\ref{fig:all_models}) exhibit fluctuations on small to intermediate length scales, although the global geometry of the viscous ejecta is fairly close to spherical symmetry.

\section{Results}
\label{sec-results}

\begin{figure*}
\includegraphics[width=0.93\textwidth]{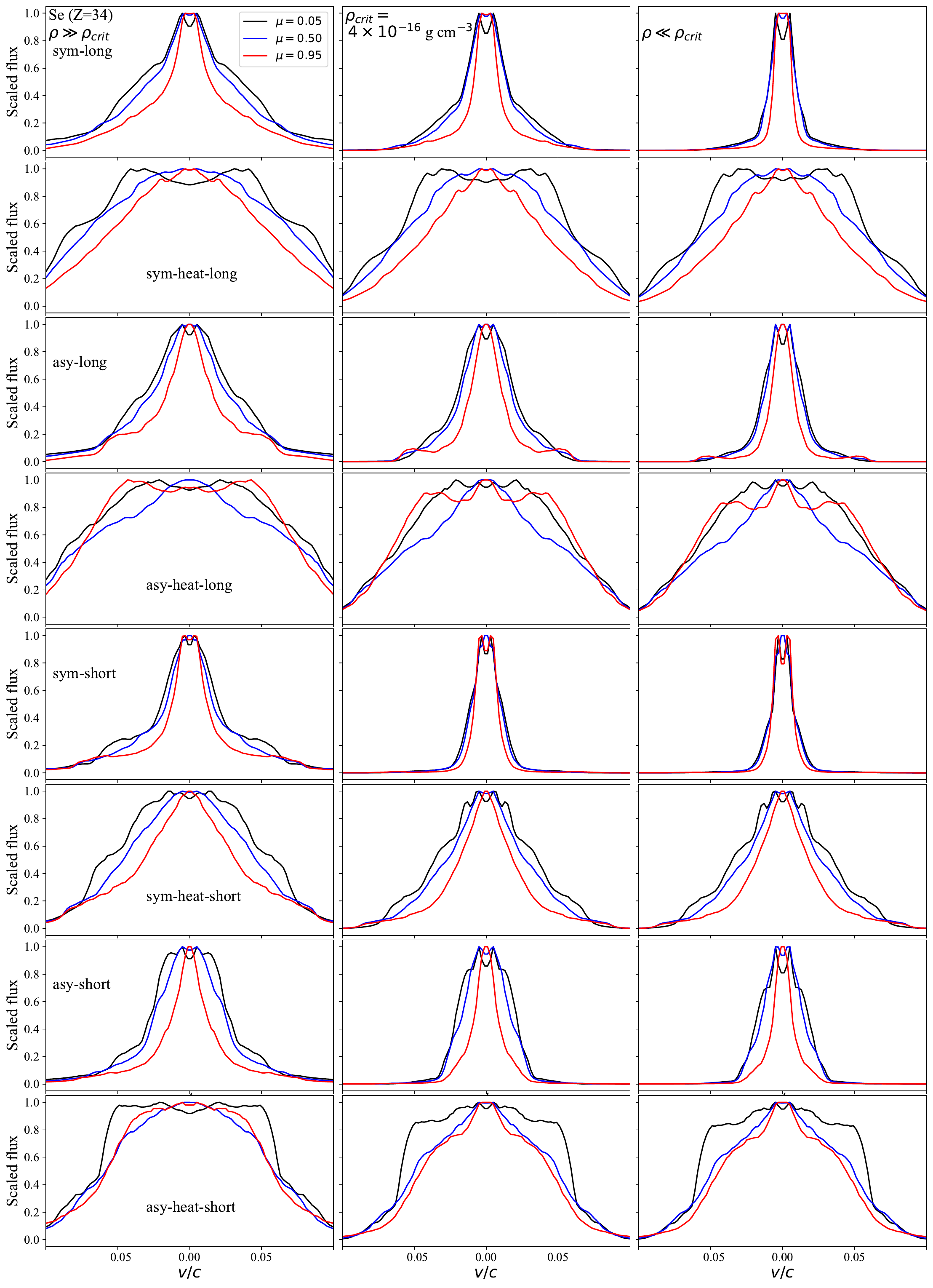}\\[-4mm]
\caption{Computed line profiles for Se ($Z=34$) for models sym-long, 
sym-heat-long,
asy-long,
asy-heat-long,
sym-short,
sym-heat-short,
asy-short and 
asy-heat-short
(ordered from top to bottom row). The left and right panels show profiles computed assuming the high- and low-density limits (i.e. $\rho \gg \rho_{\rm crit}$ and $\rho \ll \rho_{\rm crit}$), respectively, while the middle panels show results adopting $\rho_{\rm crit} = 4.2 \times 10^{-16}$~g~cm$^{-3}$. In each panel, three profiles are shown corresponding to distant observers with lines of sight defined by direction cosines ($\mu = \cos \theta$) measured relative to the polar ($z$) axis: $\mu = 0.05$ (near equatorial), $0.5$ and $0.95$ (near polar). All profiles are shown in velocity space ($v/c$) and are arbitrarily normalized for ease of comparison.}
\label{fig:Se_profiles}
\end{figure*}

\begin{figure*}
\includegraphics[width=0.93\textwidth]{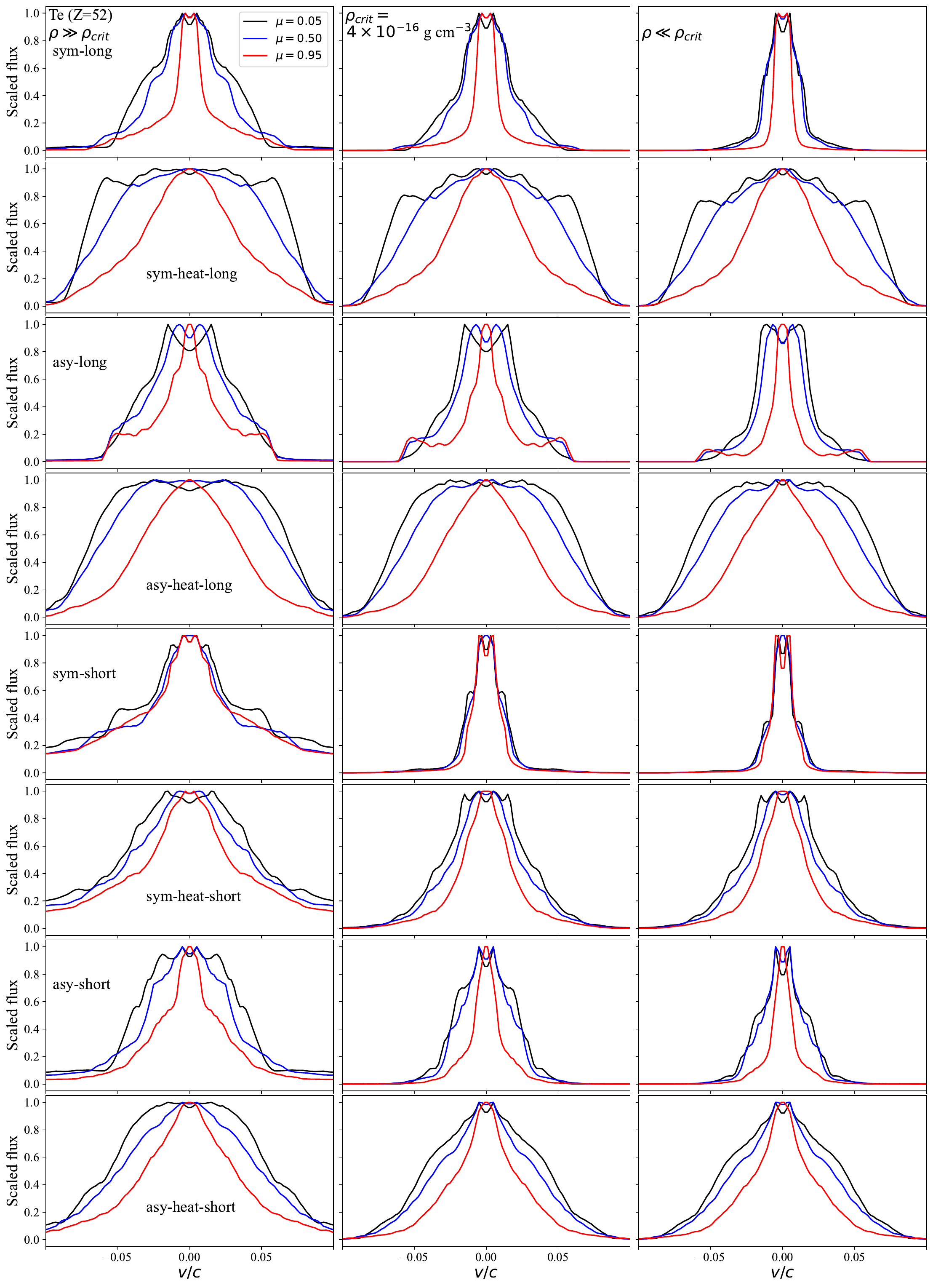}
\caption{As Fig.~\ref{fig:Se_profiles} but showing profiles calculated for Te ($Z=52$).}
\label{fig:Te_profiles}
\end{figure*}

\begin{figure*}
\includegraphics[width=0.93\textwidth]{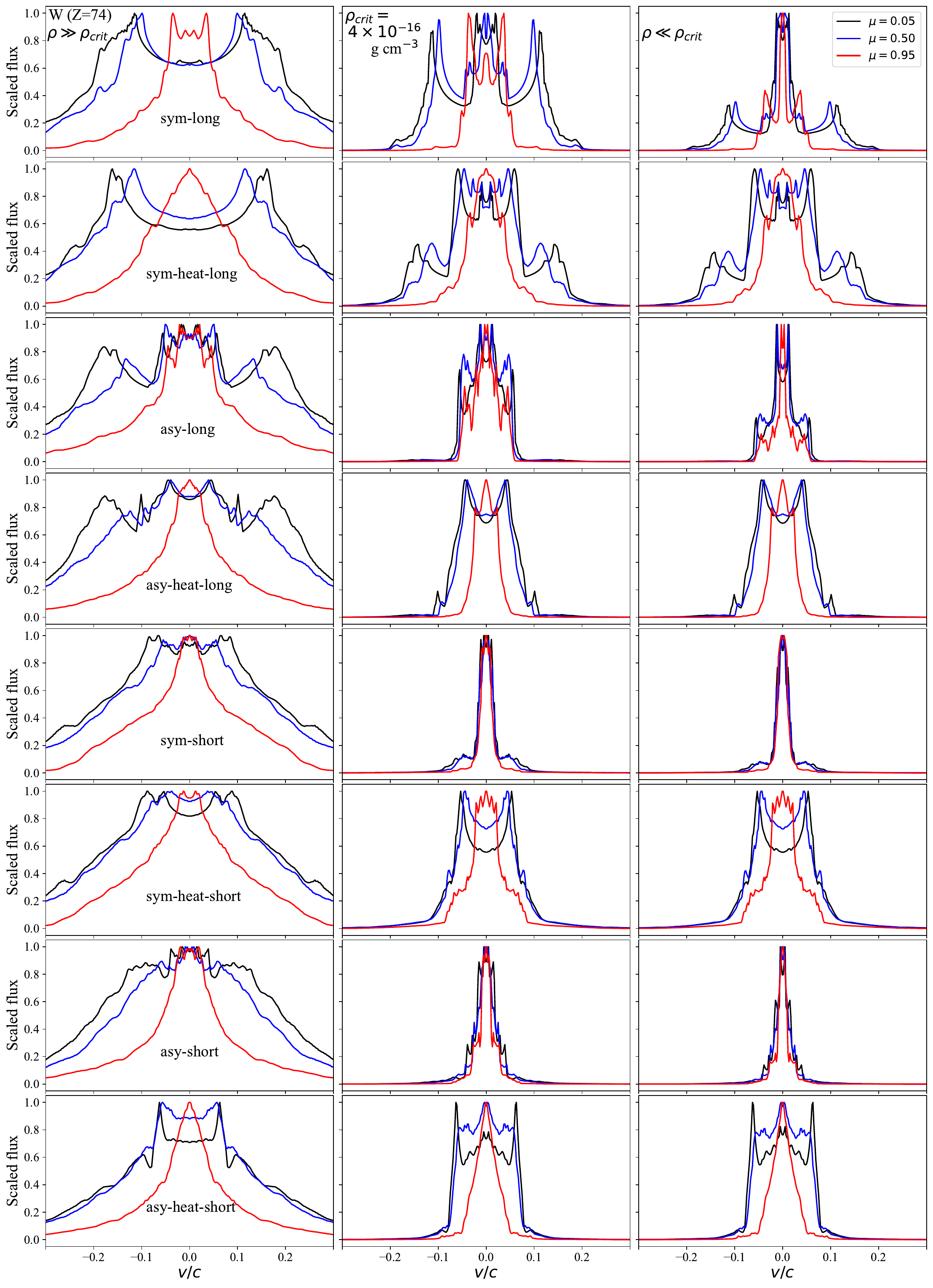}
\caption{As Fig.~\ref{fig:Se_profiles} but showing profiles calculated for W ($Z=74$). Note the wider velocity range compared to Figs.~\ref{fig:Se_profiles} and \ref{fig:Te_profiles}.}
\label{fig:W_profiles}
\end{figure*}

For each of the models considered, Figs.~\ref{fig:Se_profiles}, \ref{fig:Te_profiles} and \ref{fig:W_profiles} show computed line profiles for three observer orientations\footnote{{Note that our method (see Section~\ref{sec-method}) does not involve any angular binning or integration over observer orientation. Our calculations should therefore be appropriate for a distant observer whose line-of-sight lies at an angle of $\theta = \cos^{-1} \mu$ to the  $z$-axis.}} ({direction cosines of} $\mu = 0.05$ [{corresponding to} polar angle $\theta = $ {$ \cos^{-1} \mu =$}  $ 87.3^\circ$,  i.e. near-equatorial {orientation}], $0.5$ [$\theta = 60^\circ$, intermediate] and $0.95$ [$\theta = 18.2^\circ$, near-polar]) for Se ($Z=34$), Te ($Z=52$) and W ($Z=74$), respectively. 
or each element in each model the profile shapes are computed to illustrate both the high- and low-density limits (i.e. assuming, respectively,  $\rho \gg \rho_{\rm crit}$ or $\rho \ll \rho_{\rm crit}$) in Eqn.~\ref{eqn:particle_summation}. We also show calculations using the full expression in Eqn.~\ref{eqn:particle_summation}, for a characteristic choice of $\rho_{\rm crit} = 4.2 \times 10^{-16}$~g~cm$^{-3}$. This value is selected to correspond to the particular example of the [\ion{Te}{iii}] 2.1~$\mu$m transition (see Appendix~A), for which a proposed identification in AT2017gfo was made by \citet{hotokezaka23}, and which may also be present in AT2023vfi \citep{levan2024, gillanders2025}. As noted in Appendix~A, the value of the critical density will vary from transition to transition but, within our framework, this will primarily amount to differences in the epoch at which the profile transitions from the high- to low-density limit.

\begin{figure*}
\includegraphics[width=1.5\columnwidth]{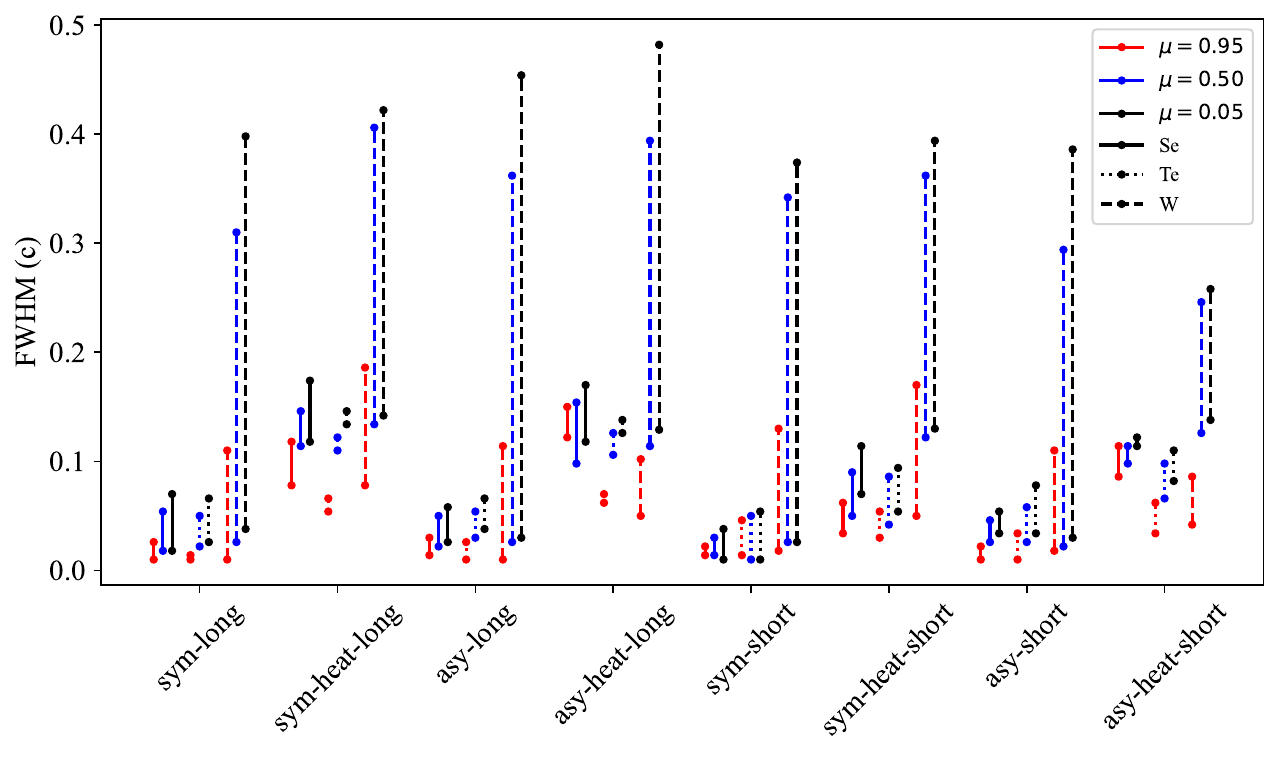}
\caption{For each model we plot FWHM values (from Table~\ref{tab:fwhm}) for Se, Te and W (left to right for each model, drawn with solid, dotted and dashed lines respectively), each shown for three observer orientations (near-polar [$\mu = 0.95$, red], intermediate [$\mu = 0.50$, blue], and near-equatorial [$\mu = 0.05$, black]). In each case, a vertical line is drawn that shows the range between values obtained adopting the low- and high-density limit in our calculation.}
\label{fig:FWHM}
\end{figure*}

To further quantify our results, we tabulate full-width at half maximum (FWHM) values obtained from selected profiles in Table~\ref{tab:fwhm} and illustrate how this measure of line width varies in Fig.~\ref{fig:FWHM}. As discussed below, however, we caution that our profile shapes are complex, meaning that a simple FWHM measurement does not reliably capture the extent of line wings in many cases.
In particular, the W profiles sometimes show multiple distinct peaks and can include both very narrow cores and extended wings. In such cases, the value tabled in Table~\ref{tab:fwhm} refers to the narrow core and is given in parenthesis, and we refer the reader to Fig.~\ref{fig:W_profiles} for the full profile shape.

\begin{table*}
\caption{For each model we give the measured FWHM for near-polar, intermediate and near-equatorial lines of sight ($\mu = 0.95$, $0.50$ and $0.05$) for our high- and low-density limit calculations. We note that none of our profiles are Gaussian in shape and that in many cases the models predict complicated wing shapes that mean the full extent of the profile is not well-captured by a FWHM value alone. In some cases, the W profiles show multiply-peaked profiles with very narrow cores. In such cases, the FWHM presented here is given in parenthesis ( ... ) and refers to the inner core of the profile, but we stress that these values are not representative of the full velocity range of the profiles.
Values are derived to an accuracy of approximately $\pm 0.004c$.}
\label{tab:fwhm}
\begin{tabular}{cccccccc} \hline
 Model & $\mu$ & \multicolumn{2}{c}{Se FWHM ($c$)} & \multicolumn{2}{c}{Te FWHM ($c$)}
 & \multicolumn{2}{c}{W FWHM ($c$)}\\
 &  & $\rho \gg \rho_{\rm crit}$  &  $\rho \ll \rho_{\rm crit}$ & $\rho \gg \rho_{\rm crit}$  &  $\rho \ll \rho_{\rm crit}$ & $\rho \gg \rho_{\rm crit}$  &  $\rho \ll \rho_{\rm crit}$ \\ \hline
sym-long & $0.95$ & 0.026& 0.010& 0.014 & 0.010& 0.110 & (0.010)\\
 & $0.50$ & 0.054 & 0.018 & 0.050 & 0.022& 0.310& (0.026)\\
 & $0.05$ & 0.070 & 0.018 & 0.066 & 0.026& 0.398& (0.038)\\
sym-heat-long & $0.95$ & 0.118 & 0.078 & 0.066 & 0.054 & 0.186 & 0.078\\ 
 & $0.50$ & 0.146 & 0.114 & 0.122 & 0.110 & 0.406 & (0.134)\\
 & $0.05$ & 0.174 & 0.118 & 0.146 & 0.134 & 0.422 & (0.142) \\ \hline
asy-long & $0.95$ & 0.030 & 0.014 & 0.026 & 0.010 & 0.114 & (0.010)\\
 & $0.50$ & 0.050 & 0.022 & 0.054 & 0.030 & 0.362 & (0.026)\\
 & $0.05$ & 0.058 & 0.026 & 0.066 & 0.038 & 0.454 & (0.030)\\
asy-heat-long & $0.95$ & 0.150 & 0.122 & 0.070 & 0.062 & 0.102 & 0.050\\
 & $0.50$ & 0.154 & 0.098 & 0.126 & 0.106 & 0.394 & 0.114\\
& $0.05$ & 0.170 & 0.118 & 0.138 & 0.126 & 0.482 & 0.129\\ \hline
sym-short & $0.95$ & 0.022 & 0.014 & 0.046 & 0.014 & 0.130 & 0.018\\ 
 & $0.50$ & 0.030 & 0.014 & 0.050 & 0.010 & 0.342 & 0.026\\
 & $0.05$ & 0.038 & 0.010 & 0.054 & 0.010 & 0.374 & 0.026\\
sym-heat-short & $0.95$ & 0.062 & 0.034 & 0.054 & 0.030 & 0.170 & 0.050\\
 & $0.50$ & 0.090 & 0.050 & 0.086 & 0.042 & 0.362 & 0.122\\
& $0.05$ & 0.114 & 0.070 & 0.094 & 0.054 & 0.394 & 0.130\\ \hline
asy-short  & $0.95$ & 0.022 & 0.010 & 0.034& 0.010& 0.110 & 0.018\\ 
 & $0.50$ & 0.046& 0.026 & 0.058 & 0.026 & 0.294 & (0.022)\\
 & $0.05$ & 0.054& 0.034 & 0.078 & 0.034 & 0.386 & (0.030)\\ 
asy-heat-short  & $0.95$ & 0.114 & 0.086 & 0.062& 0.034& 0.086 & 0.042\\ 
 & $0.50$ & 0.114& 0.098 & 0.098 & 0.066 & 0.246 & 0.126\\
 & $0.05$ & 0.122& 0.114 & 0.110 & 0.082 & 0.258 & 0.138\\ \hline 
\end{tabular}
\end{table*}

Our calculations predict clear variations in profile shapes, owing to the different ejecta structures. 
We start by noting three trends that emerge systematically. First, near-polar lines of sight ($\mu = 0.95$) tend to produce narrower profiles than near-equatorial orientations ($\mu = 0.05$), with intermediate orientations lying somewhere between (see Figs.~\ref{fig:Se_profiles} -- \ref{fig:FWHM}). This is a consequence of the fact that the element distributions (Fig.~\ref{fig:all_models}) have some degree of equatorial concentration, meaning that projected line-of-sight velocities from the emitting region are largest for $\mu$ close to zero. This effect manifests for almost all models and elements we consider (see Fig.~\ref{fig:FWHM}), but the scale varies from case to case and we note that some exceptions emerge (e.g. Se in the asy-heat-long model; see Fig.~\ref{fig:Se_profiles}).
Also, we suspect that if emission lines were identified from elements that are predominantly produced in high $Y_e$ regions, i.e. along the poles, those might exhibit opposite trends.

Second, line profiles computed in the high-density limit are broader than those obtained in the low-density limit. This follows from Eqn~\ref{eqn:particle_summation}, where $\rho \gg \rho_{\rm crit}$ leads to flux contributions that are proportional to element masses ($M_{\rm elem}$), while $\rho \ll \rho_{\rm crit}$ leads to contributions that are additionally weighted by the local density. Since the mass density is always centrally concentrated (see left panels of Fig.~\ref{fig:all_models}), this means profiles obtained from the low-density limit are narrower. In principle, this trend could emerge as a tendency for observed profiles to narrow over time (as densities drop in the expanding ejecta). However, we note that for the characteristic $\rho_{\rm crit} = 4.2 \times 10^{-16}$~g~cm$^{-3}$ that we consider, the profiles we obtain at one month post explosion are generally already very close to the low-density case (compare middle to right columns in Figs.~\ref{fig:Se_profiles}, \ref{fig:Te_profiles} and \ref{fig:W_profiles}). This implies that any noticeable profile evolution would require either significantly earlier phases, at which point our optically thin assumption would be less well justified, or transitions with significantly lower critical densities.

Third, we find that including energy due to r-process heating in the hydrodynamical models significantly impacts the predicted profile shapes. Specifically, the profiles computed for our `heat' models tend to be systematically broader (see Fig.~\ref{fig:FWHM}). The difference is most striking for profiles that are formed mainly in the inner ejecta -- i.e., in our calculation, the fractional change in line width is typically larger for Se and Te than W, and larger in the low-density limit than high-density cases. This follows from the different structures of the models (Fig.~\ref{fig:all_models}) where it is apparent that the heating effect inflates the inner ejecta. 

The complicated ejecta morphology means that the line profile shapes are complex. Our profiles for the lower mass elements considered (Se and Te) tend to consist of a narrow core (velocities of $\sim0.01$ -- $0.02c$ for models that neglect r-process heating, rising to $\sim 0.03c$ -- $0.05c$ when this heating is included) combined with more extended wing emission, which may extend to $\sim0.05c$ --$0.1c$ but is generally fairly weak. The narrow shape for these elements results from the circumstance that lighter r-process elements are produced mainly (and with the highest densities) in the low-velocity BH-torus winds (cf. Fig.~\ref{fig:ej_components1}). In some cases, the Se and Te profiles show double peaks and/or broad shoulders, and the diversity of profile shapes is even more striking for W (Fig.~\ref{fig:W_profiles}). 
{The relatively small amount of W at low velocities means that the W profiles lack strong emission in the line core and are more dominated by high-velocity contributions. This leads to {\it normalized} profiles that are significantly broader than for the lighter elements.}
They often include distinct extended wings and/or separated peaks on scales of $\sim 0.1c$, most prominently for near-equatorial observation angles, as a result of the equatorial concentration of dynamical ejecta. Particularly for calculations in the low-density limit, we caution that the W profiles in particular are sensitive to the model predictions for any small quantities of low-$Y_{e}$ material that are present in the innermost ejecta -- such material is responsible for the narrow cores (and associated complex morphology) in these cases.

\section{Discussion}
\label{sec-discuss}

Our calculations demonstrate that the multi-dimensional structure of the ejecta predicted by neutron star merger models can significantly affect the shapes of optically-thin emission line profiles, which can be relevant for the interpretation of observations in several ways. 
 
We found that line profiles for our representative first-, second- and third-peak r-process elements are narrower for near-polar observations. This offers a potential test of models by examining variations in profile widths for samples of kilonovae that span a range of observer inclinations, once such observational datasets are available. While the orientation effect is systematic, the absolute widths vary significantly between models, meaning that measurements of line widths for any single object would not directly disentangle orientation from other model parameters. Nevertheless, particularly if orientation can be estimated by other means \citep[as for e.g. AT2017gfo, see][]{Mooley2018}, profile width provides a constraint on other merger properties. Or, if comparing multiple observations of events with known masses, differences in profile width could be used to obtain an ordering by inclination.
It is also apparent that line shapes could help to identify from which element a specific feature emerged if several candidates have been deemed possible based on wavelength and luminosity considerations. For instance, Fig.~\ref{fig:FWHM} shows that one would expect an increased line width for W as compared to Se and Te.

We have found that the ejecta structure can lead to profile shapes that are clearly non-Gaussian and even involve multiple peaks.
Our W profiles can show peaks separated from the line centre by as much as $0.1c$, while profiles for Se and Te can show broad shoulders and structured profiles extending to velocities of a few percent of $c$. These complexities are certainly on a scale large enough to be relevant to JWST observations. For example, \cite{gillanders2025} discuss multiple-component fits to the spectra of AT2023vfi that are motivated by substructure in the observed 2.1$\mu$m line on scales of $\sim 0.04c$. 
{The non-Gaussianity of our profile functions is not surprising:  although widely used as a convenient function for line fitting and/or characterizing fluxes and widths, exactly Gaussian line profile shapes are not generally expected for late-time emission lines from explosive transients \cite[see e.g.][]{jerkstrand17}. However, more examples of high quality late-time observations and modelling that accounts for ionization/temperature variations will be needed to accurately assess consistency between models are data. We note that if observations were to establish that line profiles are significantly less structured and/or less orientation dependent than models predict, this would likely be evidence that the ejecta are more spherically symmetric than current models \citep[see e.g.][]{sneppen23}.}

Comparing our results with the W profile calculations from \cite{mccann25}, we note that the trend they discussed is still present in our calculations for equivalent models. In particular, \cite{mccann25} presented profile calculations in the low-density limit from spherically averaged models that did not consider the impact of r-process heating. They found that models involving a long-lived neutron star remnant led to broader W profiles. Our low-density calculations for models that do not consider r-process heating do show a similar behaviour, namely that the W profiles for the long-lived models have stronger/broader wing emission than the short-lived models (see Fig.~\ref{fig:W_profiles}). However, comparisons of our findings to \cite{mccann25} also demonstrate the importance of the multi-dimensional structures: the width of the line wings is significantly affected by observer orientation, and a very narrow line core is present in all our W profiles from models that did not include r-process heating. Furthermore, this comparison also highlights the impact of including r-process heating when simulating the ejecta: in all calculations, the line core is significantly broader for models in which this effect is taken into account. 

The differences between our model calculations illustrate how ejecta kinematics and morphology could be constrained by profile widths and shapes. Although complex, such differences can be connected back to differences in the merger parameters. For example, a signature may be imprinted by the binary mass ratio, as our models predict a systematically larger amount of heavy elements in low-velocity BH-torus ejecta from asymmetric mergers (compared to symmetric) and, correspondingly, less significant wing broadening. This is apparent, for example, in our W profile calculations for which the symmetric mergers show somewhat more prominent wings than the corresponding asymmetric examples (Fig.~\ref{fig:W_profiles}). 
One may also speculate about the extreme case (not considered in this study) of a very long-lived (or even stable) NS remnant that is believed to deposit its rotational energy into the ejecta over timescales of several seconds via magnetar spin down \citep[e.g.][]{Duncan1992a, Metzger2008f}. Based on the tendencies found in this study, such a scenario could lead to very extended or complex line wings for the majority of r-process elements, because the low-velocity ejecta may be significantly accelerated into some ring-like or extended structure. Ultimately, constraints on the ejected element distribution may even help to constrain the equation of state for high-density matter \citep[see e.g.][]{sneppen2026}, although establishing such connections would require a much broader survey of models than considered here.

Of particular note for future modelling is our finding that including r-process heating during the explosion dynamics can significantly alter (i.e. broaden) the line profiles, particularly for features in the inner ejecta (Se and Te, in our examples). It is therefore important that this effect should be included if they are to be used as input for late-phase spectral analyses.

\section{Outlook}
\label{sec-outlook}

In summary, future line-profile observations may offer the possibility not only to infer the total mass of a given element but also to constrain the geometric distribution and, if line observations are available for several elements, the compositional stratification within the ejecta. Due to their sensitivity to all of the ejected material -- and at late times particularly to high-density material typically located at low velocities -- line-profile observations could provide important complementary information to what can be learned from the early photospheric KN signal, which is mainly determined by the relatively fast ejecta.

The investigation presented here can, however, only be viewed as an exploratory proof of principle. Likely the most severe limitation is that it is expected that almost all observed features will involve blends of multiple transitions \citep[][]{pognan23, pognan25a, pognan2026, jerkstrand25b, gillanders24, gillanders2025, hotokezaka21, hotokezaka22}, meaning that full models accounting for emission complexes with realistic calculations of relative strengths would be needed for accurate inferences to be drawn from comparison to data. Full calculations must also consider residual opacity, most importantly from permitted lines \citep[see e.g. ][]{pognan23}, which will affect the emergent spectrum. 
Furthermore, our simple assumptions about the plasma conditions, particularly those of uniform ionization and temperature (Section~\ref{sec-method}) will not be accurate \citep[see e.g. temperature profiles obtained by][]{pognan23, pognan25a, jerkstrand25a, pognan2026}, and will
impact profile shapes. 
For example, it is plausible that the mean degree of ionization or temperature could vary systematically with velocity \cite[see e.g. ][]{pognan23} or polar angle in the ejecta. Compared to the calculations here, this could lead to systematic differences in the relative strength of emission in line cores compared to line wings which may vary from case to case, depending on the ion/transition in question.
An additional caveat to our results is that the symmetries in the models we consider (rotational, and reflection in $xy$-plane) lead to profiles that are symmetric about line centre. Full 3D models can break this symmetry which ultimately adds further complexity to the range of possible shapes that has not been explored here (see Appendix B). Apart from the dimensionality, the line shapes may also be sensitive to other approximations or assumptions adopted in the hydrodynamic simulation models to make them computationally tractable -- such as an effective turbulent viscosity, missing magnetic fields, absence of an ultra-relativistic jet outflow component, approximate treatment of general relativity and neutrino transport -- of which the impact needs to be explored in more detail in future works.

Nevertheless, intrinsic profile shapes are an important element of understanding the spectra. Even when line blending is significant, the intrinsic shapes are relevant to understanding the structure of pseudo-continua that might be formed by lines and our results clearly support the value of gathering high-quality observational data that has the potential to constrain profiles for a range of species. 

\section*{Acknowledgments}

This work is funded/co-funded by the European Union (ERC, HEAVYMETAL, 101071865). GL and GMP 
acknowledge support by
  the European Research Council (ERC) under the European Union’s Horizon 2020 research and innovation programme
 (ERC Advanced Grant KILONOVA No. 885281). ZX  is funded by the European Union under the ERC Grant (NeuTrAE, No. 101165138). Views and opinions expressed are, however, those of the authors only and do not necessarily reflect those of the European Union or the European Research
Council. Neither the European Union nor the granting authority can be held responsible for them.
CEC is funded by the European Union’s Horizon Europe
research and innovation programme under the Marie Skłodowska-Curie grant
agreement No.~101152610. OJ, ZX, AB, GL, GMP, and LJS acknowledge
support by the Deutsche Forschungsgemeinschaft (DFG,
German Research Foundation) through Project No. ID
279384907—SFB 1245 (Subprojects No. B06 and
No. B07). We thank the anonymous referee for their comments on the paper.

\section*{Data Availability}

The data underlying this article will be shared on reasonable request to the corresponding author.



\bibliographystyle{mnras}
\bibliography{biblio} 




\appendix

\section{Analytic form for the photon emission coefficient}

\begin{figure}
\includegraphics[width=0.95\columnwidth]{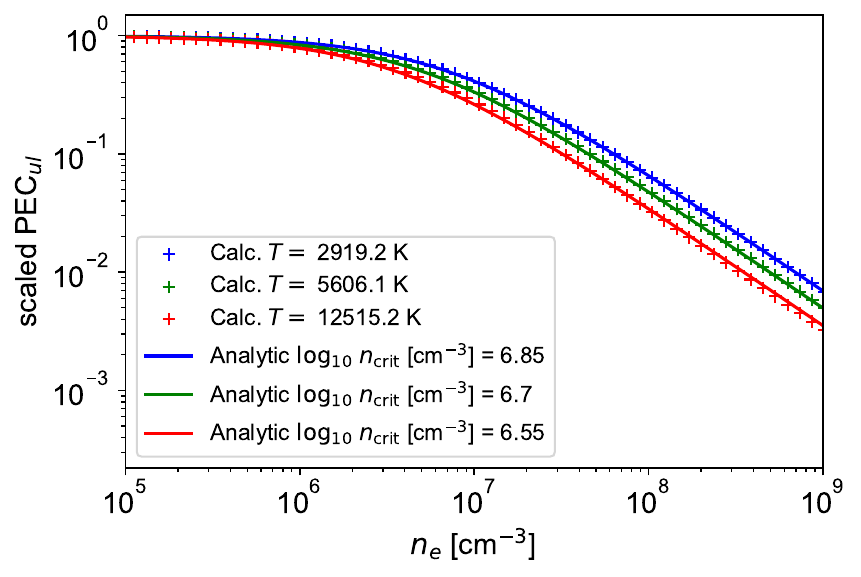}
\caption{Photon emission coefficients (${\rm PEC}_{ul}$) 
for the [\ion{Te}{iii}] 2.1~$\mu$m transition at three temperatures.
Symbols indicate the computed values from \citet{mulholland24}. 
Solid lines indicate the analytic form 
(Eqn.~\ref{eqn:critical}) 
for critical values of the electron density ($n_{\rm crit}$) as indicated. In all cases, ${\rm PEC}_{ul}$ is scaled to 1.0 in the low density limit.
}
\label{fig:pec}
\end{figure}

As discussed in Section~\ref{sec:pec}, we assume that the density dependence of the photon emission coefficient can be approximated by the simple form given in Eqn.~\ref{eqn:critical}, which describes emissivity for a collisionally excited transition from an optically thin plasma. Fig.~\ref{fig:pec} illustrates the accuracy of this simple analytic form by comparing to values of
${\rm PEC}_{ul}$ computed for the [\ion{Te}{iii}] 2.1~$\mu$m transition by \cite{mulholland24} for three illustrative temperatures. As expected, the shape of the curve is well matched by the analytic form (Eqn.~\ref{eqn:critical}) across a wide range of electron density. 
The value of the critical density ($n_{\rm crit}$) 
is temperature dependent but varies by only a factor of a few across the temperature range of interest for kilonova modelling (e.g. by about a factor of two across the range $T \sim 3,000$ -- $13,000$~K shown in Fig.~\ref{fig:pec}).

For the calculations throughout this paper, we cast the critical density in terms of a critical mass density for which we adopt $\rho_{\rm crit} = 4.2 \times 10^{-16}$~g~cm$^{-3}$. Assuming a mean ion mass of $100$~a.m.u. and a doubly-ionized charge state, this corresponds to a free-electron density of $\sim 5.3 \times 10^6$~cm$^{-3}$ ($\log_{10} n_e [\mbox{cm}^{-3} ] = 6.7$), which falls around the mid-point of the temperature range considered for the [\ion{Te}{iii}] 2.1~$\mu$m transition (see Fig.~\ref{fig:pec}). 
Of course, the critical density depends on the transition in question, meaning that a full treatment for any specific line should adopt the value specific to that case. However, the simple approach of using one representative value is sufficient here since Eqn.~\ref{eqn:critical} will capture the high- and low-density limits and, in the context of homogously expanding kilonova ejecta, changing the value of the critical density is equivalent to shifting the epochs across which the evolution between these limits takes place (i.e. the profile for a transition with a higher critical density would reach the low-density limit sooner).

\section{Model without reflection symmetry}
As described in Section~\ref{sec:models}, all calculations presented in Section~\ref{sec-results} have reflection symmetry in the $xy$-plane. Together with the assumed rotational symmetry about the $z$-axis this means the profile shapes we calculated in the optically thin limit are all symmetric about line centre. However, if the symmetry between positive and negative hemispheres is broken, it is to be expected that asymmetries will appear between the red and blue sides of the profile. Hemispheric asymmetries may arise due to stochastic fluctuations in the fluid flow \citep[e.g.][]{Collins24}, which in turn may be triggered by hydrodynamic instabilities that can operate during or after the merger, such as the Kelvin-Helmholtz instability \citep[e.g.][]{Oechslin2007} acting in the shear layers between both stars or the convective instability \cite[e.g.][]{Fernandez2013b, Just2015a} caused by viscous heating in the remnant disk\footnote{We note, however, that the viscosity treatment in the hydrodynamic models used here is merely an approximate way of describing the effects that actually would result from the magneto-rotational instability \citep[e.g.][]{Siegel2017b}.}.

\begin{figure*}
\includegraphics[height=0.25\textwidth]{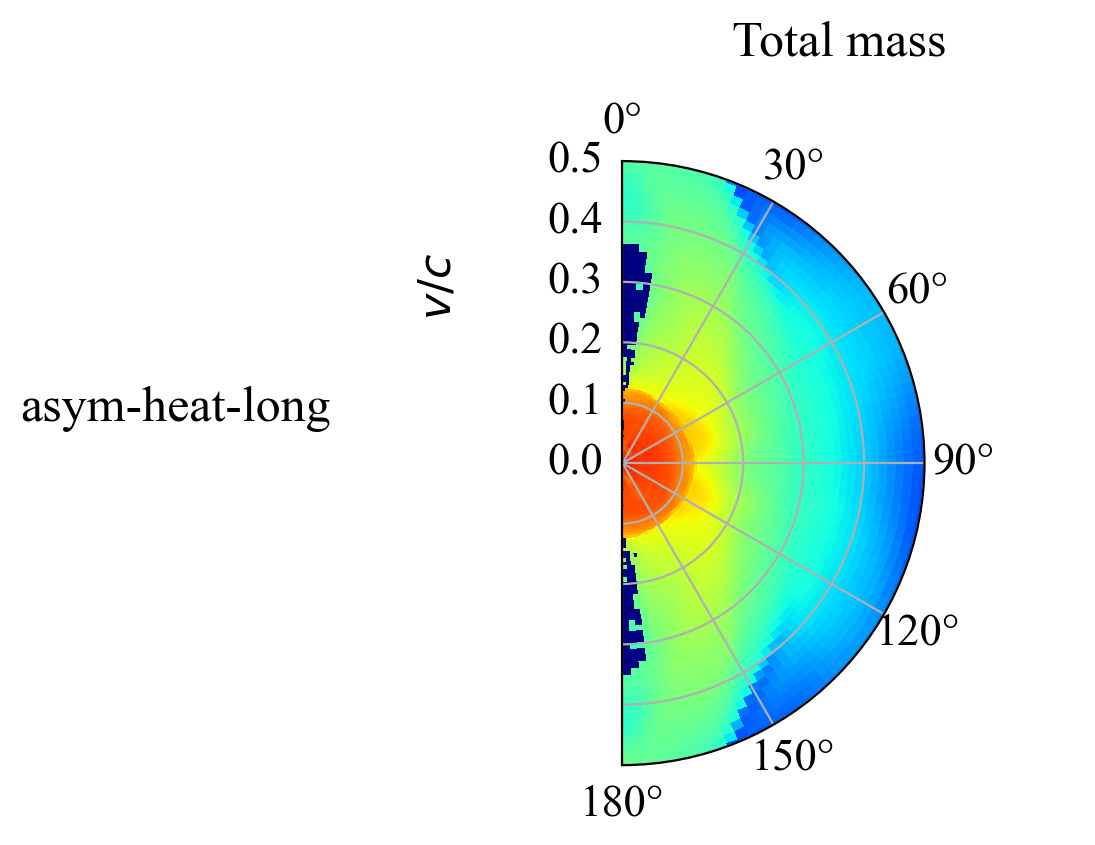}
\includegraphics[height=0.25\textwidth]{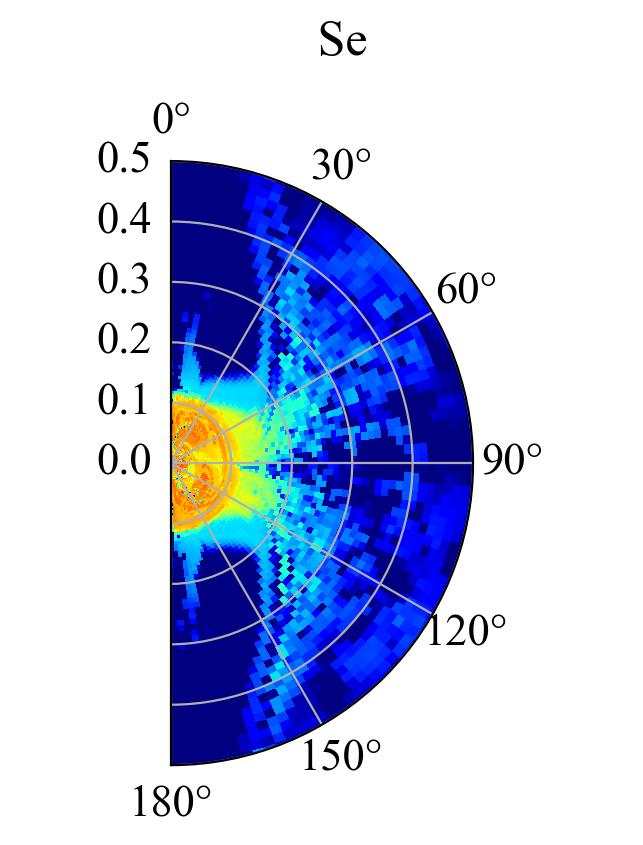}
\includegraphics[height=0.25\textwidth]{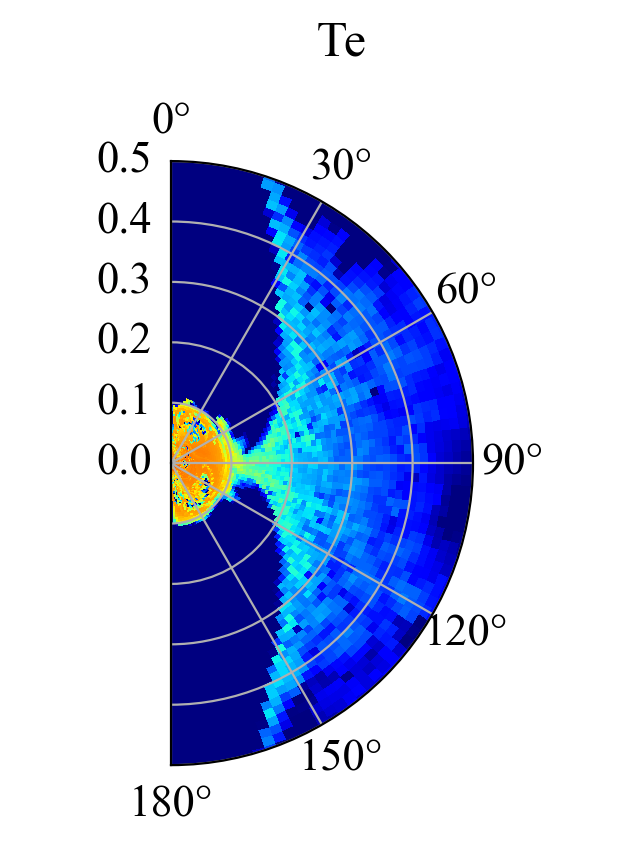}
\includegraphics[height=0.25\textwidth]{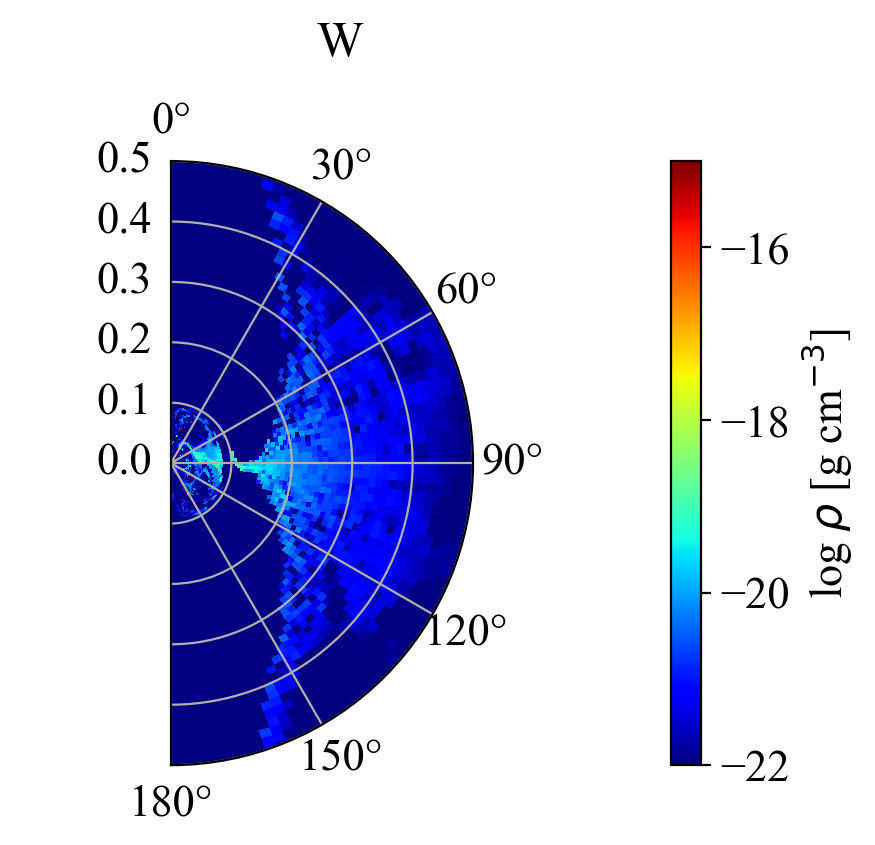}
\includegraphics[height=0.225\textwidth]{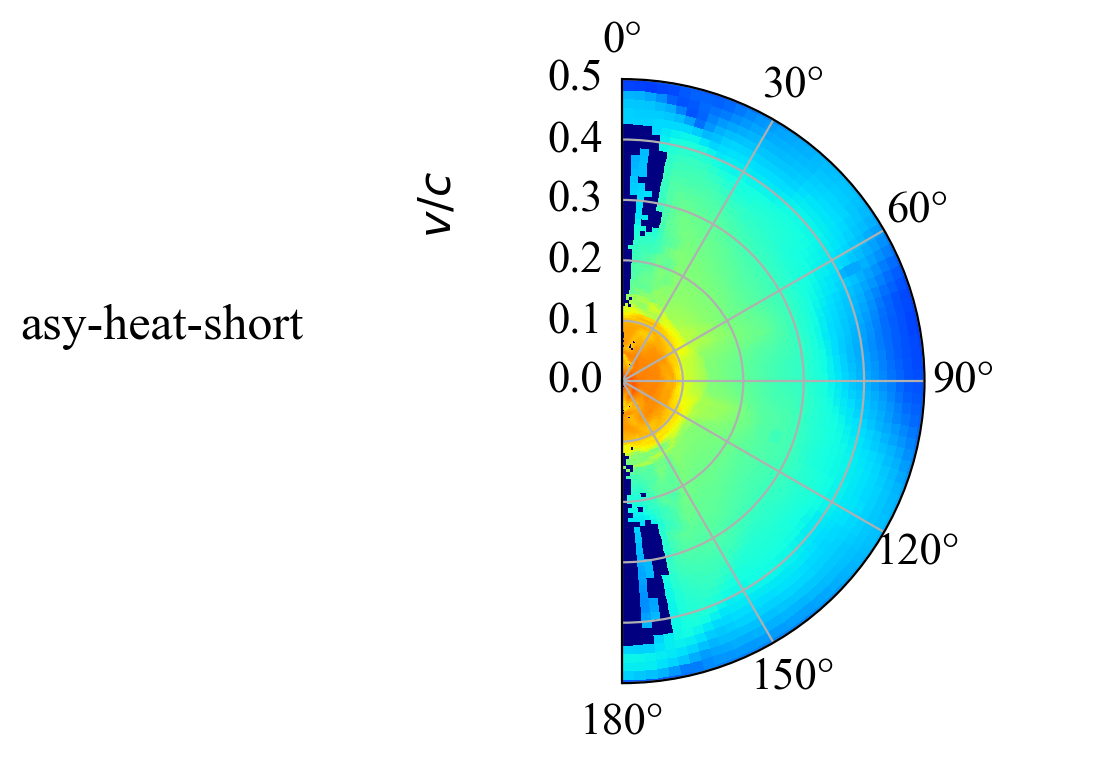}
\includegraphics[height=0.225\textwidth]{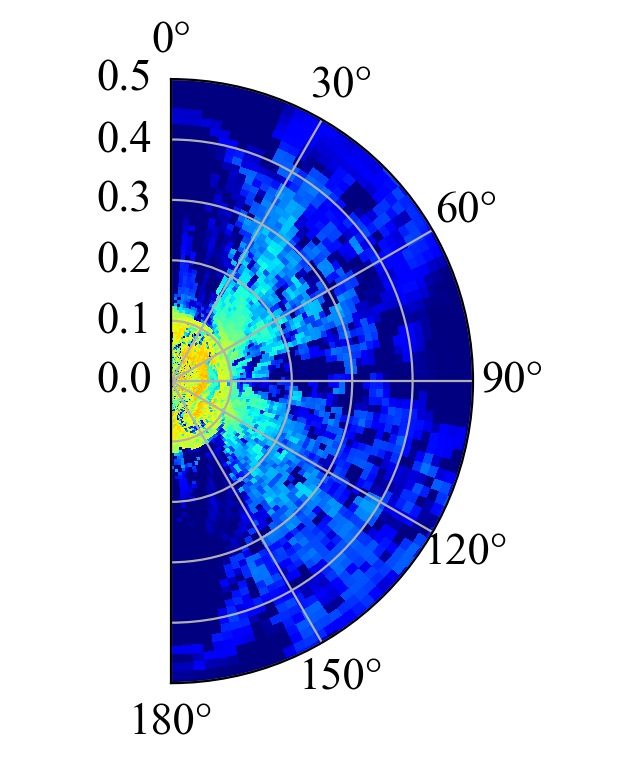}
\includegraphics[height=0.225\textwidth]{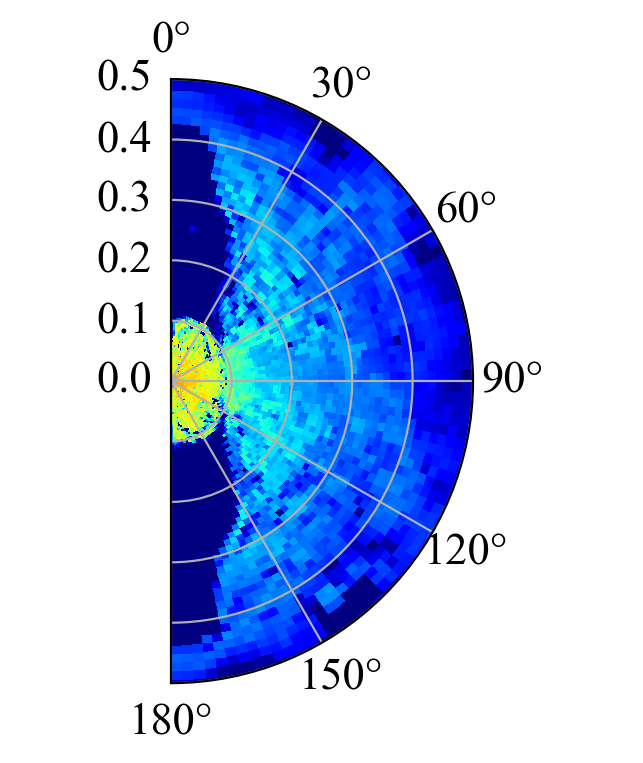}
\includegraphics[height=0.225\textwidth]{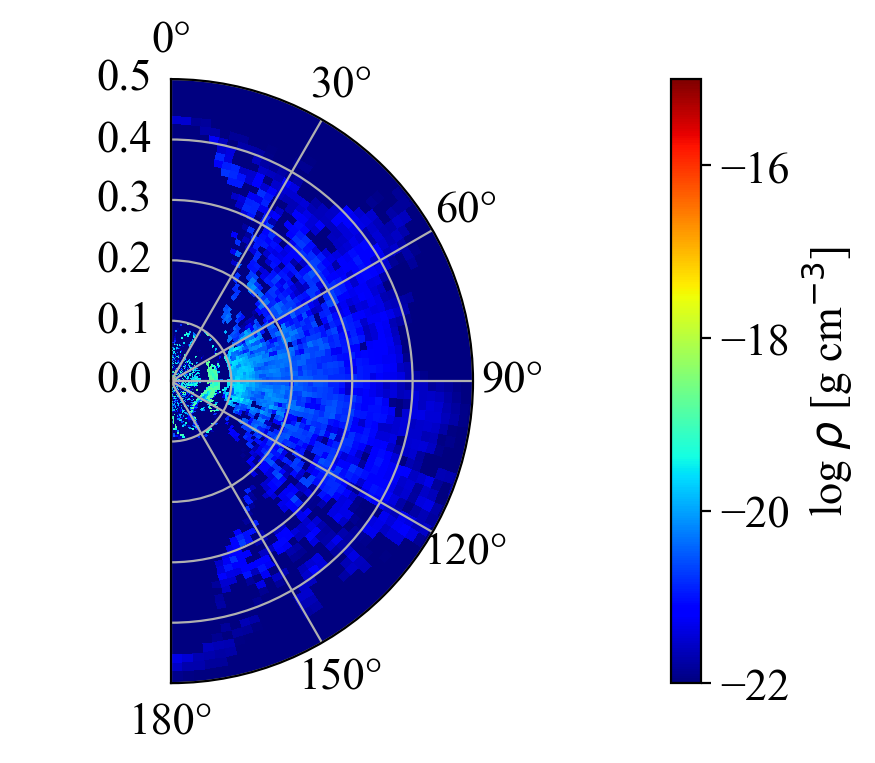}
\caption{Similar to Fig.~\ref{fig:all_models}, but showing mass density ($\rho$, left) and elemental mass densities for Se ($Z=34$, 2nd column), Te ($Z=52$, 3rd column) and W ($Z=74$, right column) using the full angular range (i.e. both hemispheres) for model asy-heat-long (upper) and asy-heat-short (lower) models. Compositions are taken at 1 month ($= 30$~days) post merger and densities are plotted in velocity space ($v/c$).}
\label{fig:fullmodels}
\end{figure*}

\begin{figure*}
\includegraphics[width=0.9\textwidth]{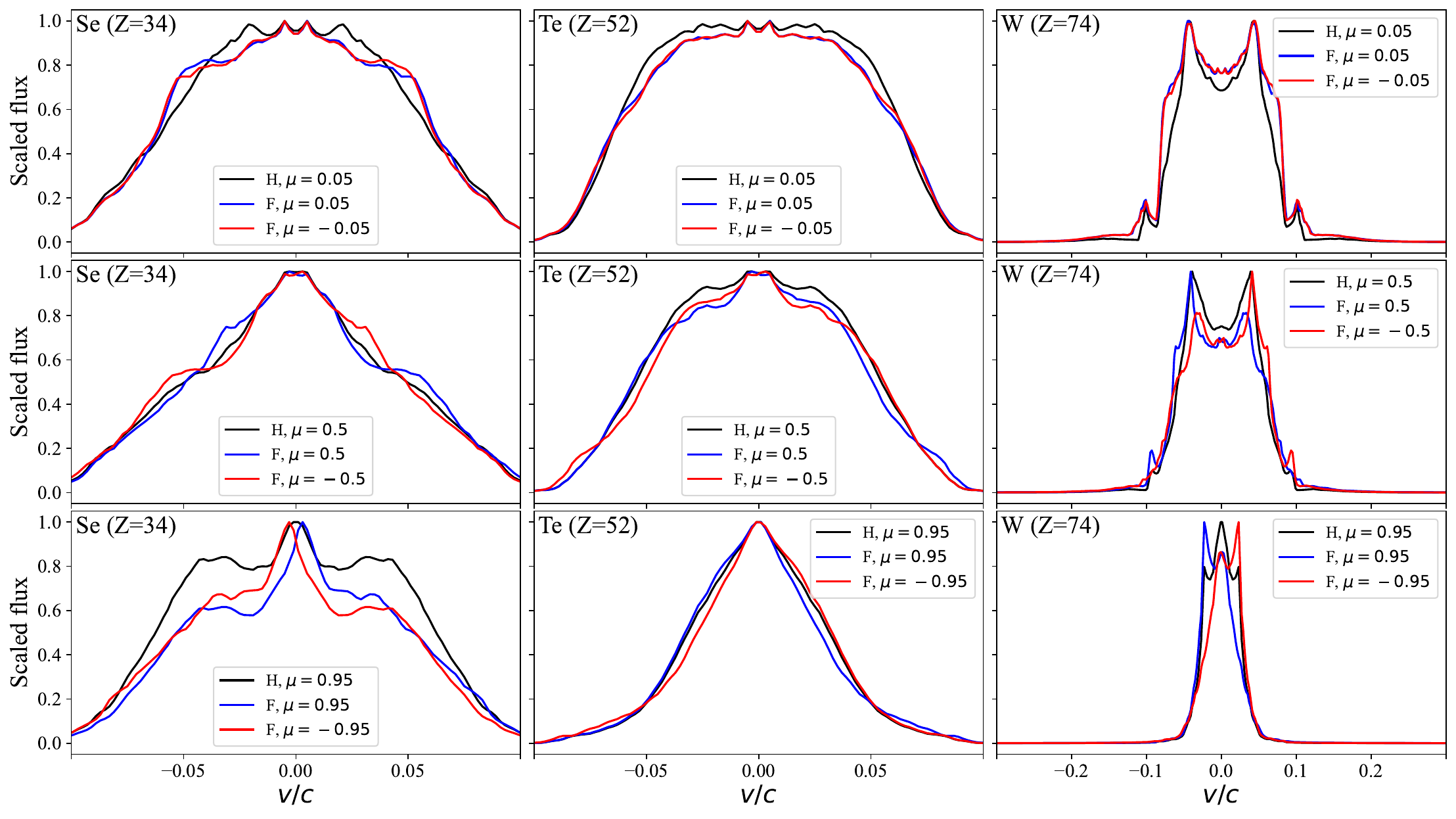}
\caption{Computed line profiles for Se ($Z=34$), Te ($Z=52$) and W ($Z=74$) for the asy-heat-long model adopting the $\rho \ll \rho_{\rm crit}$ limit. In each panel the profile obtained from the version of this model used throughout the rest of this manuscript (H, i.e. in which the negative $z$ hemisphere is assumed to be an $xy$-plane reflection of the positive hemisphere) is shown in black. Blue and red show profiles computed using the full model (F) for observer orientations in both hemispheres that have the same angle to the polar axis. From top to bottom we show cases for near-equatorial, intermediate and near-polar orientations. All profiles are shown in velocity space ($v/c$) and are arbitrarily normalized for ease of comparison.
}
\label{fig:fullmodel_profiles_asylong}
\end{figure*}

\begin{figure*}
\includegraphics[width=0.9\textwidth]{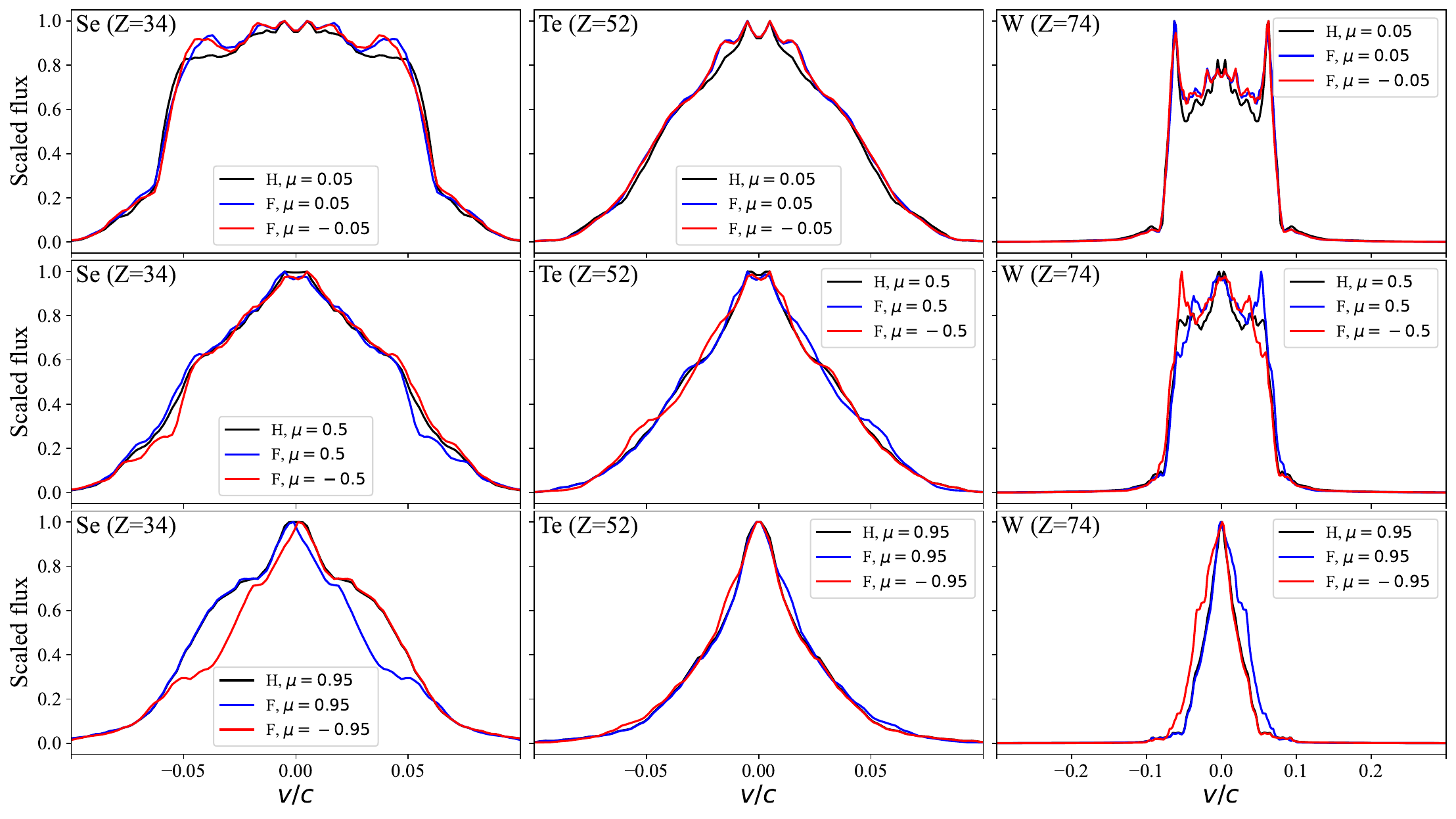}
\caption{As Fig.~\ref{fig:fullmodel_profiles_asylong} but showing the asy-heat-short model.
}
\label{fig:fullmodel_profiles_asyshort}
\end{figure*}

To give an indication of the possible impact of asymmetries between the positive and negative hemispheres we have therefore recomputed profiles for two such models (asy-heat-long and asy-heat-short) using input data for the full angular range. The structure of the full models that we use are shown in Fig.~\ref{fig:fullmodels}, and computed profiles for near-polar, intermediate and near-equatorial lines of sight for Se, Te and W are shown in Figs.~\ref{fig:fullmodel_profiles_asylong} and \ref{fig:fullmodel_profiles_asyshort} (computed adopting the low-density limit, $\rho \ll \rho_{\rm crit}$).

As Fig.~\ref{fig:fullmodels} shows, the gross structure of the negative $z$ hemisphere is very similar to the positive hemisphere, but there are some small asymmetries, and these are somewhat more prominent in the short-lived model (lower panels).
The impact of the hemisphere asymmetry on the profiles is most prominent for near-polar lines of sight (see Fig.~\ref{fig:fullmodels}) since it is for polar observers that the opposite hemispheres correspond to the approaching and receding sides of the ejecta. In general, the impact on the models considered here is modest but does lead to asymmetries in the line wings on a scale of $\sim 0.01$ to $0.02$c.

For near-equatorial observers the differences between the calculations are even smaller. However, we stress that for these orientations breaking the blue/red symmetry of the profile would depend on departures from rotational symmetry about the $z$ axis (rather than $xy$-reflection symmetry) which is not captured in any of the models we consider here. Thus we must defer to future work any further investigation of the range of possible asymmetric profiles that might be generated from fully 3D ejecta models.


\bsp	
\label{lastpage}
\end{document}